\documentclass[journal,12pt,onecolumn,draftclsnofoot,]{IEEEtran}

\ifCLASSINFOpdf
  \usepackage[pdftex]{graphicx}
  \graphicspath{{./images/}}
  \DeclareGraphicsExtensions{.pdf,.jpeg,.png}
\else
\fi

\usepackage{amsmath}
\usepackage{amssymb}
\usepackage{commath}
\usepackage{cases}
\usepackage{cite}
\usepackage{xcolor}
\makeatletter
\let\NAT@parse\undefined
\makeatother
\usepackage{hyperref}
\hypersetup{linkbordercolor=green}
\usepackage{hhline}
\usepackage{tikz}
\usepackage{pgfplotstable}
\usetikzlibrary{fit}
\usetikzlibrary{positioning,automata}

\makeatletter
\def\bstctlcite{\@ifnextchar[{\@bstctlcite}{\@bstctlcite[@auxout]}}
\def\@bstctlcite[#1]#2{\@bsphack
  \@for\@citeb:=#2\do{%
    \edef\@citeb{\expandafter\@firstofone\@citeb}%
    \if@filesw\immediate\write\csname #1\endcsname{\string\citation{\@citeb}}\fi}%
  \@esphack}
\makeatother

\usepackage[ruled, lined, linesnumbered]{algorithm2e}
\usepackage{threeparttable}

\ifCLASSOPTIONcompsoc
  \usepackage[caption=false,font=normalsize,labelfont=sf,textfont=sf]{subfig}
\else
  \usepackage[caption=false,font=footnotesize]{subfig}
\fi

\newcommand{\B}[1]{\mathbf{#1}}
\newcommand{\pr}[1]{\left(#1\right)}
\newcommand{\br}[1]{\left[#1\right]}
\newcommand{\bbr}[1]{\left\{#1\right\}}
\newcommand{\nr}[1]{\left\|#1\right\|}

\newtheorem{theorem}{Theorem}
\newtheorem{remark}{Remark}

\newtheorem{assumption}{Assumption}

\definecolor{bluePerso}{RGB}{0, 102, 204}
\definecolor{greenPerso}{RGB}{0, 153, 0}
\definecolor{guppiegreen}{rgb}{0.0, 1.0, 0.5}

\makeatletter
\newcommand{\removelatexerror}{\let\@latex@error\@gobble}
\makeatother

\makeatletter
\newcommand{\thickhline}{%
    \noalign {\ifnum 0=`}\fi \hrule height 1pt
    \futurelet \reserved@a \@xhline
}

\begin{document}
\bstctlcite{IEEEexample:BSTcontrol}

\title{Split-and-augmented Gibbs sampler -- \\Application to large-scale inference problems}

\author{Maxime~Vono, Nicolas~Dobigeon and Pierre~Chainais%

\thanks{Maxime Vono and Nicolas Dobigeon are with the University of Toulouse, IRIT/INP-ENSEEIHT, CNRS, 2 rue Charles Camichel, BP 7122, 31071 Toulouse cedex 7, France (e-mail: Maxime.Vono@irit.fr, Nicolas.Dobigeon@enseeiht.fr).}
\thanks{Pierre Chainais is with Univ. Lille, CNRS, Centrale Lille, UMR 9189 - CRIStAL - Centre de Recherche en Informatique Signal et Automatique de Lille, F-59000 Lille, France (e-mail: Pierre.Chainais@centralelille.fr).}}

\maketitle

\begin{abstract}
This paper derives two new optimization-driven Monte Carlo
algorithms inspired from variable splitting and data augmentation.
In particular, the formulation of one of the proposed approaches is closely related to the alternating direction method of multipliers (ADMM) main steps.
The proposed framework enables to derive faster and more efficient sampling schemes than the current state-of-the-art methods and can embed the latter.
By sampling efficiently the parameter to infer as well as the hyperparameters of the problem, the generated samples can be used to approximate Bayesian estimators of the parameters to infer.
Additionally, the proposed approach brings confidence intervals at a low cost contrary to optimization methods. Simulations on two often-studied signal processing problems illustrate the performance of the two proposed samplers.
All results are compared to those obtained by recent state-of-the-art optimization and MCMC algorithms used to solve these problems.
\end{abstract}

\begin{IEEEkeywords}
Bayesian inference, data augmentation, high-dimensional problems, Markov chain Monte Carlo, variable splitting.
\end{IEEEkeywords}

\IEEEpeerreviewmaketitle

\section{Introduction}
\label{sec:Introduction}
\IEEEPARstart{N}{umerous} machine learning, signal and image processing problems involve the estimation of a hidden object of interest $\B{x} \in \mathbb{R}^N$ based on (noisy) observations $\B{y} \in \mathbb{R}^M$.
This unknown object of interest can stand for parameters of a given model in machine learning \cite{Hastie2001} or may represent a signal or image to be recovered within an inverse problem.
With the increasing amount and variety of available data, solving such inference problems in high dimension becomes challenging and generally relies on sophisticated computational inference methods.
Those methods are mainly based on stochastic simulation and variational optimization which are two powerful tools to perform inference in complex models \cite{Pereyra2016A}.
An important class of stochastic simulation techniques is the family of the Markov chain Monte Carlo (MCMC) methods \cite{Robert2004}. Within a Bayesian inference framework, MCMC algorithms have the great advantage of providing a comprehensive description of the posterior distribution of the parameter $\B{x}$ to be inferred. Contrary to optimization techniques which generally provide a point estimate, this description permits the subsequent derivation of credibility intervals on the parameter $\B{x}$.
Nonetheless, note that optimization algorithms can also bring confidence information when the log-likelihood is supposed differentiable by relying on the theory of large samples \cite{Kendall46}.
These confidence measures are particulary important for inference problems where very few observations are available (e.g. in biology \cite{Hwang2004}, physics \cite{VonToussaint2011} or astrophysics \cite{Loredo1992}) or when one is interested in extreme events (e.g. in hydrology \cite{Reis2005} or cosmology \cite{Trotta2008}).
For instance, MCMC methods have been recently used to conduct Bayesian inference on gravitational waves \cite{Veitch2015}.
However, contrary to optimization techniques, MCMC methods may suffer from their high computational cost which can be prohibitive for high-dimensional problems.
To overcome this limitation, a few attempts have been made to derive optimization-driven Monte Carlo methods. The Hamiltonian Monte Carlo method \cite{Duane1987}, also referred to as hybrid Monte Carlo, is an archetypal example of the successful use of variational analysis concepts (i.e., gradients) to facilitate the exploration of the target distribution. More recently, Pereyra \cite{Pereyra2016B} proposed an innovative combination of convex optimization and MCMC algorithms. Capitalizing on the advantages of proximal splitting recently popularized to solve large-scale inference problems \cite{Combettes2005,Figueiredo2003a,Daubechies2004,Elad2006,Hale2008,Bioucas-Dias2007}, the proximal Monte Carlo method allows high-dimensional log-concave distributions to be sampled. For instance, this algorithm has been successfully used to conduct antisparse coding \cite{Elvira2017} and has been significantly improved in \cite{Durmus2018}.

Concurrently, variable splitting methods, developed at least 70 years ago \cite{Courant1943}, have been recently and extensively used to solve large-scale inference problems of the form
\begin{equation}
\label{eq:optimProblem}
\underset{\B{x}}{\arg \min}\ f(\B{x}) + g(\B{x}),
\end{equation}
where $f$ commonly refers to a data fitting term and $g$ stands for some regularization function which is often nonsmooth and/or even nonconvex.
The main idea of those methods consists in splitting the variable of interest $\B{x}$ into a pair of variables $\B{x}$ and $\B{z}$ and then solving the counterpart minimization problem
\begin{align}
\label{eq:optimProblem_varSplitting}
  \begin{split}
&\underset{\B{x},\B{z}}{\arg \min}\ f(\B{x}) + g(\B{z}), \\
&\text{subject to } \B{x} = \B{z}.
  \end{split}
\end{align}
The equality constraint ensures that solving \eqref{eq:optimProblem_varSplitting} is equivalent to solve the initial problem \eqref{eq:optimProblem}. Exploiting the variable splitting idea, the alternating direction method of multipliers (ADMM) \cite{Boyd2011}, firstly introduced in \cite{Gabay1976,Glowinski1975}, has proven to be considerably faster than fast iterative thresholding-shrinkage algorithms (FISTA) \cite{Beck2009} for solving high-dimensional inverse problems in signal/image processing \cite{Afonso2010,Afonso2011}.
This increase in speed comes from the fact that ADMM uses a second-order information of the data fidelity term whereas ISTA or FISTA essentially only takes into account gradient information.
The efficiency of ADMM makes it stand as a reference method in high-dimensional signal processing problems such as those encountered in hyperspectral imaging  \cite{Thouvenin2016,Halimi2017}.
This paper, in the same spirit as \cite{Pereyra2016B}, attempts to reconcile optimization and Bayesian inference by proposing two new optimization-driven MCMC algorithms that do not sample directly from the usual target distribution
\begin{equation}
\label{eq:targetDistribution}
\pi(\B{x}) \propto \exp\br{-f(\B{x})-g(\B{x})},
\end{equation}
which is assumed to be proper in the sequel.
The first one is only based on the idea of variable splitting and considers a joint probability distribution $p(\B{x},\B{z})$ which tends towards (\ref{eq:targetDistribution}) in a limiting case. The main purpose is to work with two simpler distributions $\propto\exp\br{-f(\B{x})}$ and $\exp\br{-g(\B{z})}$ separately.
A similar scheme was recently and independently proposed by \cite{Rendell2018} in order to distribute Monte Carlo methods on possibly multiple machines. The second proposed approach goes one step further by introducing an auxiliary variable $\B{u} \in \mathbb{R}^N$ within a data augmentation scheme.
The main rationales behind the proposed approaches are threefold.
Firstly, fully Bayesian approaches allow other parameters (e.g. nuisance or regularization hyperparameters) to be jointly estimated with the parameter of interest $\B{x}$, avoiding their empirical and painful hand-tuning.
Secondly, as emphasized above, samples generated by MCMC algorithms can be used to build confidence intervals on the estimated parameters contrary to optimization techniques that only provide a point estimate.
Finally, variable splitting and data augmentation within the proposed approach pave the way towards faster and more efficient samplers.

To this purpose, Section \ref{sec:Model} introduces the hierarchical Bayesian models associated to the proposed approaches.
In particular, the main ingredients, namely variable splitting and data augmentation, are presented.
Section \ref{sec:Inference} derives the two resulting optimization-driven MCMC algorithms called SP (splitting) and SPA (splitting \& augmentation).
In particular, a parallel between ADMM and the proposed SPA algorithm is drawn.
Section \ref{sec:Bayesian_inference_problems} considers two often-studied inference problems encountered in signal processing that require to sample respectively from high-dimensional Gaussian and log-concave probability distributions.
Section \ref{sec:Experiments} illustrates the performance of the proposed algorithms on these inference problems.
Finally, Section \ref{sec:Conclusion} draws concluding remarks.

\section{Model}
\label{sec:Model}
This section introduces the proposed approach which aims at using variable splitting and data augmentation to accelerate and simplify the solving of large scale Bayesian inference problems.
The main properties of the resulting joint distributions are introduced and its convergence properties towards the usual target distribution \eqref{eq:targetDistribution} are proven. Table \ref{table:list_of_symbols} summarizes the main symbols used to define the proposed models.

\begin{table}[h!]
\renewcommand{\arraystretch}{1.3}
\caption{List of symbols.}
\label{table:list_of_symbols}
\centering
\begin{tabular}{c l}
\thickhline
\bfseries Symbol & \bfseries Description \\
\hline
$\B{x},\B{z},\B{u},N$ & parameter of interest, auxiliary variables\\
& and their dimension \\
$\B{y},M$ & observation vector and its dimension\\
$f,g$ & data fitting term and regularization function\\
$\pi$ & usual target distribution\\
$\phi_1,\phi_2$ & functions associated to the split/augmented scheme\\
$\rho,\alpha$ & parameters of the proposed approaches\\
$\mathcal{N}$ & normal distribution \\
\thickhline
\end{tabular}
\end{table}

\subsection{Variable splitting}
\label{subsec:variable_splitting}
Within an optimization framework, variable splitting aims at individually using each term $f$ and $g$ of the objective function in an optimization sub-problem.
This divide-to-conquer strategy generally yields simpler proximal operators and therefore an easier algorithm to implement \cite{Combettes2011}. Following the same intuition, in a Bayesian setting, variable splitting is expected to lead to simpler sampling steps and thereby to a more efficient sampler.
Starting from the usual target distribution \eqref{eq:targetDistribution}, the introduction of a splitting variable $\B{z} \in \mathbb{R}^N$ leads to so-called \emph{split distribution} defined by
\begin{equation}
\label{eq:jointDistribution}
\pi_{\rho} \triangleq  p(\B{x},\B{z};\rho^2) \propto \exp\br{-f(\B{x})-g(\B{z}) - \phi_1(\B{x},\B{z};\rho^2)}
\end{equation}
where $\phi_1: \mathbb{R}^{N} \times \mathbb{R}^{N} \rightarrow \mathbb{R}^+$ is a divergence such that $\pi_{\rho}$ defines a proper joint distribution and $\rho$ is a positive parameter that controls the dissimilarity between $\B{x}$ and $\B{z}$.
Interestingly, the associated conditional distributions that would be considered in a Gibbs algorithm scheme to sample according to \eqref{eq:jointDistribution} are
\begin{eqnarray}{}
  p(\B{x}|\B{z};\rho^2)\propto \exp\br{-f(\B{x}) - \phi_1(\B{x},\B{z};\rho^2)} \label{eq:SP_condDistrib_x}\\
  p(\B{z}|\B{x};\rho^2)\propto \exp\br{-g(\B{z}) - \phi_1(\B{x},\B{z};\rho^2)}. \label{eq:SP_condDistrib_z}
\end{eqnarray}
Thus, this variable splitting allows $f$ and $g$ to be dissociated with the hope that these conditional distributions be easy to sample from. 
Indeed experiments in Section \ref{sec:Experiments} will show that considering the split distribution $\pi_{\rho}$ in \eqref{eq:jointDistribution} instead of $\pi$ in \eqref{eq:targetDistribution} leads to a faster and more efficient algorithm.

It is worth noting that this variable splitting-based approach can be related to previous works \cite{Geman1992,Geman1995}, revisited and extended in \cite{Idier2001}, which also introduced auxiliary variables to split the initial objective function. However, the aforementioned works considered an exact data augmentation scheme which is not the case here, see Theorem \ref{theorem:1} below. In addition, this scheme was specifically designed for Bayesian models relying on a Gaussian likelihood function, which is much more restrictive than the target distribution \eqref{eq:targetDistribution} addressed here. Finally, the data augmentation scheme considered in \cite{Geman1992,Geman1995} may practically rise some computational difficulty since it requires closed-form expressions of the augmented prior, which could not be available in general. Nonetheless, note that both the latter and the proposed approaches can be interpreted as divide-to-conquer approaches ending up with simpler full conditional distributions.

Within a parallel setting, \cite{Rendell2018} proposed a similar variable-splitting Bayesian framework motivated by distributed computations when the likelihood function can be expressed as a sum of terms over a possibly big dataset.
Their approach can be viewed as a particular instance of the proposed approach when $f(\B{x}) = \sum_{i=1}^bf_i(\B{x})$.

\begin{figure}
  \centering
  \begin{tikzpicture}[auto, semithick, level 1/.style={sibling distance=0.9cm},edge from parent/.style={draw,latex-},font=\normalsize]

    \node[circle,draw] (x) {$\B{x}$} [grow'=up]
      child {node[circle,draw] (theta1) {$\boldsymbol{\theta}_{\B{x}}$}};

    \node[circle,draw,dashed,right=1.0cm of x,color=greenPerso] (rho) {$\rho^2$} [grow'=up];

    \node[circle,draw,color=greenPerso,right=1.0cm of theta1] (z) {$\B{z}$} [grow'=up];;

    \node[circle,draw,color=greenPerso,right=0.8cm of z] (theta2) {$\boldsymbol{\theta}_{\B{z}}$};

    \node[circle,draw,below=1cm of rho,color=bluePerso] (u) {$\B{u}$} [grow'=right];;

    \node[circle,draw,color=bluePerso,right=0.8cm of u] (theta3) {$\boldsymbol{\theta}_{\B{u}}$};;

    \node[circle,draw,dashed,color=bluePerso,below=0.8cm of u] (alpha) {$\alpha^2$};;

    \draw[-latex,draw=greenPerso] (z) -- (x);
    \draw[-latex,draw=bluePerso,fill=bluePerso] (u) -- (x);
    \draw[-latex,draw=bluePerso,fill=bluePerso] (theta3) -- (u);
    \draw[-latex,draw=bluePerso,fill=bluePerso] (alpha) -- (u);
    \draw[-latex,draw=greenPerso,fill=greenPerso] (rho) -- (x);
    \draw[-latex,draw=greenPerso,fill=greenPerso] (theta2) -- (z);

    \node[text width=1.2cm,align=center,below=0.5cm of x] (L2) {\footnotesize Variable of interest};
    \node[text width=1.5cm,align=center,below=0.5cm of theta2] (L3) {\footnotesize Variable splitting};
    \node[text width=1.5cm,align=center,below=0.5cm of theta3] (L4) {\footnotesize Data augmentation};

    \node[draw,dotted,fit=(x) (theta1) (L2)] {};
    \node[draw,dotted,fit=(z) (rho) (theta2) (L3)] {};
    \node[draw,dotted,fit=(u) (alpha) (theta3) (L4)] {};

  \end{tikzpicture}
  \caption{DAGs associated with the usual and proposed hierarchical Bayesian models. In black: DAG associated to (\ref{eq:targetDistribution}); in black and green: DAG associated to (\ref{eq:jointDistribution}); in black, green and blue: DAG associated to (\ref{eq:jointAugDistribution}). $\boldsymbol{\theta}_{\B{x}}$, $\boldsymbol{\theta}_{\B{z}}$ and $\boldsymbol{\theta}_{\B{u}}$ stand for possible additional parameters that are not discussed in this paper. (User-defined parameters appear in dashed circles).\label{fig:DAG}}
\end{figure}
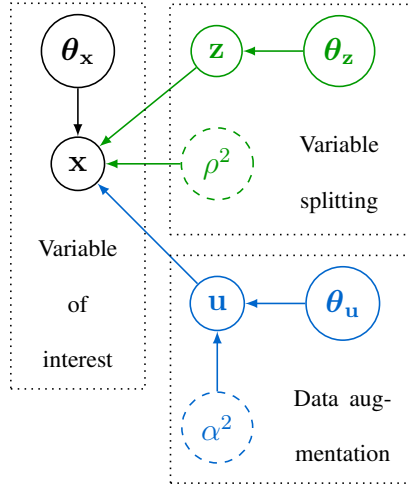

The directed acyclic graph (DAG) associated with the proposed splitting model is depicted in Fig. \ref{fig:DAG} in black and green. Note that sampling from \eqref{eq:jointDistribution} instead of \eqref{eq:targetDistribution} boils down to considering another hierarchical Bayesian model. However, to ensure the relevance of this extended model and the associated distribution \eqref{eq:jointDistribution} with respect to the inference problem underlied by the target distribution \eqref{eq:targetDistribution}, one can expect that $\phi_1$ tend to zero when $\B{z}$ is close to $\B{x}$. Thus, if $\phi_1$ is a divergence measure where the discrepancy between $\B{x}$ and $\B{z}$ is controlled by $\rho^2$, it has to satisfy the following assumption that is closely related to the equality constraint $\B{x} = \B{z}$ in variable splitting methods.

\begin{assumption}\label{assumption:1}
  Let $\B{x}$ and $\B{z}$ obeying the distribution (\ref{eq:jointDistribution}). Then, $\phi_1$ is assumed to be such that, for all $\B{x},\B{z} \in \mathbb{R}^N$,
  \setlength{\arraycolsep}{0.0em}
  \begin{eqnarray}
    \lim_{\rho^2\to0} \dfrac{\exp\br{- \phi_1(\B{x},\B{z};\rho^2)}}{\int_{\mathbb{R}^N}\exp\br{- \phi_1(\B{x},\B{z};\rho^2)}\mathrm{d}\B{z}} = \delta_{\B{x}}(\B{z}).\label{eq:assumption1}
  \end{eqnarray}
  \setlength{\arraycolsep}{5pt}
\end{assumption}

When this assumption is ensured, the usual target distribution (\ref{eq:targetDistribution}) is expected to be recovered from the marginal distribution of $\B{x}$ associated to \eqref{eq:jointDistribution} in the limiting case $\rho\to 0$.
This expectation is met when a general form of the divergence $\phi_1$ is chosen, as stated by the following theorem.
\begin{theorem}\label{theorem:1}
  Let $p_{\rho}(\B{x}) = \int_{\mathbb{R}^N} \pi_{\rho}(\B{x},\B{z})\mathrm{d}\B{z}$. Then, under Assumption \ref{assumption:1}, the following result holds
  \begin{equation}\label{eq:convergenceTV}
    \nr{\pi - p_{\rho}}_{\mathrm{TV}} \xrightarrow[\rho^2\to0]{} 0.
  \end{equation}
\end{theorem}
\begin{IEEEproof}
See Appendix \ref{appendice:proof_theorem1}.
\end{IEEEproof}
Note that the convergence in total variation implies the convergence in distribution.
Thereby, in the limiting case where $\rho^2$ tends to zero, the marginal distribution of $\B{x}$ under $\pi_{\rho}$ coincides with the usual target distribution $\pi$.
In Section \ref{sec:Bayesian_inference_problems}, the divergence $\phi_1$ will be chosen quadratic.
This choice is not a surprise since it is often used in optimization having the great advantage of being differentiable and convex.

\subsection{Data augmentation}
\label{subsec:data_augmentation}
The first proposed approach introduces the idea of variable splitting only.
It leads to a joint distribution (\ref{eq:jointDistribution}) with an additional term $\phi_1$ that controls the discrepancy between $\B{x}$ and $\B{z}$.
Since $\phi_1$ is governed by $\rho^{2}$, one might set $\rho^2$ to a small value to ensure that $\B{x}$ and $\B{z}$ will not be too far from each other (see Theorem \ref{theorem:1}). However, when sampling from \eqref{eq:jointDistribution} via its conditional distributions \eqref{eq:SP_condDistrib_x} and \eqref{eq:SP_condDistrib_z}, the smaller $\rho^2$, the higher the correlation between samples, which may deteriorate mixing properties.
One option to improve these mixing properties is to consider a data augmentation scheme. Such a strategy consists in introducing auxiliary variables within a target distribution: it is commonly used to build more efficient sampling algorithms \cite{Dyk2001} with less interactions between MCMC draws.
This issue was for instance discussed in \cite{Besag1993,Higdon1998} for the Ising and Potts models.
Along these lines, an additional variable $\B{u} \in \mathbb{R}^N$ is introduced in the previous splitting model such that
\begin{align}
  \pi_{\rho,\alpha} &\triangleq p(\B{x},\B{z},\B{u};\rho^2,\alpha^2) \\
  &\propto \exp\br{-f(\B{x})-g(\B{z}) }\label{eq:jointAugDistribution} \\
  &\times \exp\br{- \phi_1(\B{x},\B{z}-\B{u};\rho^2)-\phi_2(\B{u};\alpha^2)}
  \nonumber
\end{align}
where $\phi_2$ is a known function defined on $\mathbb{R}^N$ such that $\pi_{\rho,\alpha}$ defines a proper joint distribution and $\alpha$ is a positive parameter. The DAG associated with the so-called \emph{split-augmented} distribution (\ref{eq:jointAugDistribution}) is depicted in Fig. \ref{fig:DAG} with additional parameters  drawn in blue compared to (\ref{eq:jointDistribution}) in black \& green only.

The conditional distributions associated with the joint split-augmented distribution \eqref{eq:jointAugDistribution} are
  \begin{align}
    &p(\B{x}|\B{z},\B{u};\rho^2) \propto \exp\br{-f(\B{x}) - \phi_1(\B{x},\B{z}-\B{u};\rho^2)} \label{eq:condDistribSPA_x}\\
    &p(\B{z}|\B{x},\B{u};\rho^2) \propto \exp\br{-g(\B{z}) - \phi_1(\B{x},\B{z}-\B{u};\rho^2)} \label{eq:condDistribSPA_z}\\
    &p(\B{u}|\B{x},\B{z};\rho^2,\alpha^2) \propto \exp\br{-\phi_2(\B{u};\alpha^2)} \nonumber\\
     &\hspace{2.7cm}\times \exp\br{- \phi_1(\B{x},\B{z}-\B{u};\rho^2)}. \label{eq:condDistribSPA_u}
  \end{align}
The differences induced by data augmentation are clearly visible when comparing (\ref{eq:SP_condDistrib_x}) and (\ref{eq:SP_condDistrib_z}) with (\ref{eq:condDistribSPA_x}) and (\ref{eq:condDistribSPA_z}).
Within a Gibbs sampler scheme, the auxiliary variable $\B{u}$ could allow to decrease the correlation between $\B{x}$ and $\B{z}$ by giving an additional degree of freedom to each of the former variables. Indeed experiments in Section \ref{sec:Experiments} will show that this data augmentation scheme leads to a sampler with better mixing properties compared to the sampler associated to $\pi_{\rho}$.

However, to assess the relevance of sampling from the split-augmented (SPA) distribution $\pi_{\rho,\alpha}$ in (\ref{eq:jointAugDistribution}) instead of the split (SP) distribution $\pi_{\rho}$ in (\ref{eq:jointDistribution}), the introduction of $\B{u}$ should not alter the joint distribution (\ref{eq:jointDistribution}). Therefore $\phi_1$ and $\phi_2$ should obey the following assumption.
\begin{assumption}\label{assumption:2}
  Let $\B{x}$, $\B{z}$ and $\B{u}$ obeying the distribution (\ref{eq:jointAugDistribution}). Then, $\phi_2$ and $\phi_1$ are assumed to be such that for all $\B{x} \in \mathbb{R}^N$ and $\B{z} \in \mathbb{R}^N$,
  \setlength{\arraycolsep}{0.0em}
\begin{eqnarray}
  \int_{\mathbb{R}^N} \exp\br{- \phi_1(\B{x},\B{z}-\B{u};\rho^2)-\phi_2(\B{u};\alpha^2)}\mathrm{d}\B{u} \nonumber \\ \propto \exp\br{- \phi_1(\B{x},\B{z};\eta^2(\rho, \alpha))}.\label{eq:condition2}
\end{eqnarray}
\setlength{\arraycolsep}{5pt}
\end{assumption}
where $\eta(\rho, \alpha)$ plays the role of a parameter. 
In other words, this assumption ensures that a split distribution $\pi_{\eta}$ of the form \eqref{eq:jointDistribution} can be obtained by marginalizing the split-augmented distribution $\pi_{\rho,\alpha}$ in \eqref{eq:jointAugDistribution} with respect to $\B{u}$. For usual choices of $\phi_1$ and $\phi_2$, this assumption is satisfied, as stated in the following theorem.

\begin{theorem}\label{theorem:2}
  Let $\B{x}$, $\B{z}$ and $\B{u}$ obeying the distribution (\ref{eq:jointAugDistribution}). In the particular case where $\phi_1$ is quadratic that is
  \begin{equation}\label{eq:phi1QuadraticBis}
    \phi_1(\B{x},\B{z}-\B{u};\rho^2) = \dfrac{1}{2\rho^2}\nr{\B{x}-(\B{z}-\B{u})}^2_2
  \end{equation}
  and $\phi_2$ has the form
  \begin{equation}\label{eq:phi2Quadratic}
    \phi_2(\B{u}) = \dfrac{1}{2\alpha^2}\nr{\B{u}}^2_2,
  \end{equation}
  Assumption \ref{assumption:2} is verified with $\eta^2(\rho, \alpha) = \rho^2 + \alpha^2$ so that 
  \begin{equation}
    \phi_1(\B{x},\B{z};\eta(\rho, \alpha)) = \dfrac{1}{2\pr{\rho^2 + \alpha^2}}\nr{\B{x}-\B{z}}^2_2. \label{eq:marginalization}
  \end{equation}
\end{theorem}
\begin{IEEEproof}
  The proof consists in a straightforward marginalization within a Gaussian model, which can be easily derived, e.g., from computations similar to those in \cite[Chap. 10]{Kay1993}.
\end{IEEEproof}

In this particular case, it appears that a unique positive parameter $\eta^2(\rho,\alpha)=\rho^2 + \alpha^2$ drives the convergence of the marginal distribution of $\B{x}$ w.r.t. the split distribution $\pi_{\eta}$, that is of the same form as \eqref{eq:jointDistribution}, towards the target distribution $\pi$ in \eqref{eq:targetDistribution}. These quadratic forms of $\phi_1$ and $\phi_2$ play a special role. They are closely related to the ADMM (see Section \ref{subsec:SPA_meets_ADMM}) and will be considered in Section \ref{sec:Bayesian_inference_problems}.

Eventually, we emphasize that the proposed splitting and data augmentation methods can be easily generalized to cases where there are more than two functions $f$ and $g$, and when these functions involve distinct linear operators $\B{K}_i$ (subsampling, blur, transform...). In this case, the target distribution can be written as $\pi(\B{x}) \propto \exp\br{-\sum_i h_i(\B{K}_i\B{x})}$ where $h_i$ can stand for data fitting, regularization or other types of functions and $\B{K}_i \in \mathbb{R}^{k_i \times N}$ are arbitrary matrices, see Appendix \ref{appendice:generalized_SP_SPA}.
For this general case, Theorem \ref{theorem:1} holds and the proof can be easily derived with the same type of arguments as in Appendix \ref{appendice:proof_theorem1}.
Additionally, Assumption \ref{assumption:2} is naturally extended by considering the marginalization of each auxiliary variable $\B{u}_i$.

\section{Inference}
\label{sec:Inference}
This section presents two MCMC algorithms to infer the parameter of interest $\B{x}$ either from the split distribution $\pi_{\rho}$ in \eqref{eq:jointDistribution} or from the split-augmented distribution $\pi_{\rho,\alpha}$ in \eqref{eq:jointAugDistribution}. In particular, the proposed sampling strategies are discussed for two particular kinds of distributions frequently encountered in signal/image processing or machine learning problems. Additionally, a parallel between the proposed approach and the ADMM is drawn.

\subsection{Gibbs samplers}
\label{subsec:MCMC}

Two MCMC algorithms, denoted SP (see Algo. \ref{algo:Gibbs1}) and SPA (see Algo. \ref{algo:Gibbs2}), respectively associated with the split and split-augmented distributions (\ref{eq:jointDistribution}) and (\ref{eq:jointAugDistribution}) are presented. These algorithms are special instances of Gibbs samplers where samples are alternatively drawn according to the conditional distributions of each variable. Precisely, SP consists in sampling according to \eqref{eq:SP_condDistrib_x} and \eqref{eq:SP_condDistrib_z}, while SPA is defined by the conditional distributions \eqref{eq:condDistribSPA_x}--\eqref{eq:condDistribSPA_u}.

\begin{figure}
\removelatexerror
\begin{algorithm}[H]
    \caption{SP}
    \label{algo:Gibbs1}
     \SetKwInOut{Input}{Input}
     \Input{Functions $f$, $g$, $\phi_1$, $\phi_2$, parameter $\rho^2$, total number of iterations $T_{\mathrm{MC}}$, number of burn-in iterations $T_{\mathrm{bi}}$, initialization $\B{z}^{(0)}$}
   \For{$t \leftarrow 1$ \KwTo $T_{\mathrm{MC}}$}{%
   \textit{\% Drawing the variable of interest} \\
   Sample $\B{x}^{(t)}$ according to $p\pr{\B{x}|\B{z}^{(t-1)};\rho^2}$ (\ref{eq:SP_condDistrib_x}) \;
   \textit{\% Drawing the splitting variable} \\
   Sample $\B{z}^{(t)}$ according to according to $p\pr{\B{z}|\B{x}^{(t)};\rho^2}$ (\ref{eq:SP_condDistrib_z}) \;
   }
   \SetKwInOut{Output}{Output}
     \Output{Collection of samples $\bbr{\B{x}^{(t)},\B{z}^{(t)}}_{t = T_{\mathrm{bi}}+1}^{T_{\mathrm{MC}}}$ asymptotically distributed according to (\ref{eq:jointDistribution}).}
\end{algorithm}
\end{figure}

\begin{figure}
\removelatexerror
\begin{algorithm}[H]
    \caption{SPA}
    \label{algo:Gibbs2}
     \SetKwInOut{Input}{Input}
     \Input{Functions $f$, $g$, $\phi_1$, $\phi_2$, param. $\rho^2$, $\alpha^2$, total nb of iterations $T_{\mathrm{MC}}$, nb of burn-in iterations $T_{\mathrm{bi}}$, initialization $\B{z}^{(0)}$ \& $\B{u}^{(0)}$}
   \For{$t \leftarrow 1$ \KwTo $T_{\mathrm{MC}}$}{%
   \textit{\% Drawing the variable of interest} \\
   Sample $\B{x}^{(t)}$ according to $p\pr{\B{x}|\B{z}^{(t-1)},\B{u}^{(t-1)};\rho^2}$ (\ref{eq:condDistribSPA_x}) \;
   \textit{\% Drawing the splitting variable} \\
   Sample $\B{z}^{(t)}$ according to $p\pr{\B{z}|\B{x}^{(t)},\B{u}^{(t-1)};\rho^2}$ (\ref{eq:condDistribSPA_z}) \;
   \textit{\% Drawing the auxiliary variable} \\
   Sample $\B{u}^{(t)}$ according to $p\pr{\B{u}|\B{x}^{(t)},\B{z}^{(t)};\rho^2,\alpha^2}$ (\ref{eq:condDistribSPA_u}) \;
   }
   \SetKwInOut{Output}{Output}
     \Output{Collection of samples $\bbr{\B{x}^{(t)},\B{z}^{(t)},\B{u}^{(t)}}_{t = T_{\mathrm{bi}}+1}^{T_{\mathrm{MC}}}$ asymptotically distributed according to (\ref{eq:jointAugDistribution}).}
\end{algorithm}
\end{figure}

As suggested in Section \ref{sec:Model}, the splitting variable $\B{z}$ has been introduced to build faster and simpler simulating schemes compared to the direct sampling from \eqref{eq:targetDistribution}. If the conditional distributions of $\B{x}$ and $\B{z}$ are easy to sample from, one can apply Algo. \ref{algo:Gibbs1} or Algo. \ref{algo:Gibbs2} directly. If this is not the case despite the variable splitting strategy, one might use surrogates (e.g, Metropolis-Hastings \cite{Robert2004} or data augmentation schemes) to sample efficiently from each conditional distribution.

To be more precise, the following paragraphs discuss the efficient sampling of two particular distributions of interest, namely Gaussian and log-concave  distributions. These distributions are frequently encountered when addressing signal processing and machine learning problems, or may specifically result from the split and/or augment steps induced by the proposed schemes.

\subsubsection{Gaussian distributions} \label{subsubsec:inference_gaussian}
When $f$ stands for a data fitting term, it is often assumed to be quadratic since quadratic loss functions arise in a wide range of applicative contexts. Within a statistical framework, this choice leads to a likelihood function defined by a Gaussian probability distribution function. Following the same motivation, when $g$ is associated with a penalization, it is often supposed to be quadratic, leading to a Tikhonov regularizer and a Gaussian prior distribution, e.g., used for ridge regression. More precisely, in a general formulation, $f$ and $g$ are assumed to have the form
\begin{align}
&f(\B{x}) = \dfrac{1}{2}(\B{x}-\boldsymbol{\mu}_1)^T\B{Q}_1(\B{x}-\boldsymbol{\mu}_1) \\
&g(\B{x}) = \dfrac{1}{2}(\B{x}-\boldsymbol{\mu}_2)^T\B{Q}_2(\B{x}-\boldsymbol{\mu}_2) \label{eq:g_quadratic}.
\end{align}
where the $\B{Q}_i$ are precision matrices. Then, the corresponding target posterior distribution $\pi$ is also Gaussian
\begin{equation}
  \label{eq:gaussianPosterior}
  \pi(\B{x}) = \mathcal{N}\pr{\B{m},\B{Q}^{-1}}
\end{equation}
where
\begin{numcases}{}
\B{Q} = \B{Q}_1 + \B{Q}_2 \label{eq:Gaussian_precision}\\
\B{m} = \B{Q}^{-1}\pr{\B{Q}_1\boldsymbol{\mu}_1 + \B{Q}_2\boldsymbol{\mu}_2}.
\end{numcases}
If the two terms in (\ref{eq:Gaussian_precision}) cannot be diagonalized in the same basis (e.g., the Fourier domain), then sampling directly from (\ref{eq:gaussianPosterior}) can be computationally intensive since, e.g., it requires to invert the precision matrix $\B{Q}$.
In the very particular case where $\B{Q}_1 = \B{H}^T\B{\Omega}\B{H}$, if $\B{Q}_2$ and $\B{H}^T\B{H}$ can be diagonalized in the same basis, then direct sampling from the posterior $\pi$ can be achieved thanks to the specific auxiliary variable method proposed in \cite{Marnissi2018}, see also Section \ref{subsec:Exp_Gaussian}. If these requirements are not met, this auxiliary method cannot be implemented. Conversely, the SP and SPA strategies proposed above can be applied to dissociate the precision matrices $\B{Q}_1$ and $\B{Q}_2$ in the sampling procedure. Indeed, when the divergence $\phi_1$ is also chosen quadratic, as in Theorem \ref{theorem:1}, the conditional distributions associated to $\B{x}$ and $\B{z}$ are Gaussian with precision matrices
\begin{align}
&\B{Q_x} = \B{Q}_1 + \dfrac{1}{\rho^2}\B{I}_N \label{eq:Gaussian_precision_x}\\
&\B{Q_z} = \B{Q}_2 + \dfrac{1}{\rho^2}\B{I}_N \label{eq:Gaussian_precision_z}.
\end{align}
Again, this demonstrates the main interest of the splitting step which makes the two precision matrices appear in two separate distributions. Now, depending of the respective form of $\B{Q}_1$ and $\B{Q}_2$, one can directly sample from these conditional distributions or use surrogate methods \cite{Papandreou2011,Gilavert2015,Feron2016,Marnissi2018}, see Section \ref{subsec:BayesProblems_Gaussian} and Appendix \ref{appendice:efficientGaussianSampling_x} for more details.

\subsubsection{Non-smooth log-concave distributions} \label{subsubsec:inference_logconcave}
More generally, if the functions $f$ and $g$ are convex, then the conditional distributions of $\B{x}$ and $\B{z}$ involved in SP and SPA are log-concave. Additionally, when $f$ (resp., $g$) is non-smooth, if the divergence $\phi_1$ is convex, continuously differentiable and gradient Lipschitz, sampling from the conditional distribution associated with $\B{x}$ (resp., $\B{z}$) can be achieved thanks to the proximal Metropolis-adjusted Langevin algorithm (P-MALA) \cite{Pereyra2016B} or the proximal Moreau-Yoshida-unadjusted Langevin algorithm (P-MYULA) \cite{Durmus2018}. For instance, such cases can be encountered when $f$ results from a loss function robust against outliers, e.g., for least absolute deviation regression, or when $g$ stands for a sparsity-inducing regularization. P-MALA and P-MYULA are based on Langevin diffusion process and resort to proximal operators to build Markov chains with interesting convergence properties. The former uses an accept/reject step in order to correct the bias introduced by the considered approximations.
On the other hand, the latter removes this Metropolis-Hasting correction step to accelerate the sampling and gives bounds on the convergence rate of the Markov chains.\\

To summarize, instead of sampling from (\ref{eq:targetDistribution}) thanks to the direct use of the previously discussed state-of-the-art MCMC algorithms, the proposed approach aims at preparing and simplifying their implementations to sample according to the conditional distributions associated with the split and split-augmented distributions. In other words, adapted efficient methods are applied to conduct specific and simpler sampling steps where $f$ and $g$ are dissociated.
Thereby, the proposed methodology does not aim at totally replacing efficient existing MCMC algorithms but can be interpreted as a ``divide-and-conquer'' approach that simplifies the task of each sampler to make the whole sampling algorithm faster.

\subsection{When SPA meets ADMM}
\label{subsec:SPA_meets_ADMM}
This ``divide-and-conquer'' idea is also at the heart of ADMM which allows simpler minimization sub-problems to be considered during the optimization process.
This relation with the proposed approach is strengthened by another similarity between SPA and ADMM.
More precisely, let consider the particular case where $\phi_1$ and $\phi_2$ have the forms (\ref{eq:phi1QuadraticBis}) and (\ref{eq:phi2Quadratic}) respectively (in agreement with the assumptions required by Theorems \ref{theorem:1} and \ref{theorem:2}), and assume that $f$ and $g$ are convex. Then, computing the MAP estimates instead of sampling in each step of Algo. \ref{algo:Gibbs2} boils down to the ADMM \cite{Boyd2011}, see Algo. \ref{algo:ADMM}.  Within this optimization framework, $\B{z}$ corresponds to the splitting variable, $\B{u}$ stands for the scaled Lagrange multiplier and $\rho^{-2}$ for the penalty parameter.

\begin{figure}
\removelatexerror
\begin{algorithm}[H]
    \caption{ADMM (scaled version)}
    \label{algo:ADMM}
     \SetKwInOut{Input}{Input}
     \Input{Functions $f$, $g$, penalty parameter $\rho^2$, initialization $t \leftarrow 0$ and $\B{z}^{(0)}$,$\B{u}^{(0)}$}
   \While{stopping criterion not satisfied}{%
   \textit{\% Minimization w.r.t. $\B{x}$} \\
   $\B{x}^{(t)} \in \arg \min_{\B{x}} -\log p\pr{\B{x}|\B{z}^{(t-1)},\B{u}^{(t-1)};\rho^2}$\;
   \textit{\% Minimization w.r.t. $\B{z}$} \\
   $\B{z}^{(t)} \in \arg \min_{\B{z}} -\log p\pr{\B{z}|\B{x}^{(t)},\B{u}^{(t-1)};\rho^2}$\;
   \textit{\% Dual ascent} \\
   $\B{u}^{(t)} = \B{u}^{(t-1)} + \B{x}^{(t)} - \B{z}^{(t)}$ \;
   \textit{\% Updating iterations counter} \\
   $ t \leftarrow t + 1$ \;
   }
   \SetKwInOut{Output}{Output}
     \Output{Approximate solution of the optimization problem $\hat{\B{x}}$.}
\end{algorithm}
\end{figure}

The ADMM is known to be an efficient optimization algorithm for high-dimensional problems. It simplifies the optimization problem by considering several simpler optimization sub-problems where advanced optimization tools and methods (e.g., proximal operators) can be embedded and applied efficiently.
Additionally, it covers a large panel of optimization problems and can be generalized to the case where more than two functions $f$ and $g$ are considered. As noticed in the previous section, this generalization property also applies to the proposed SP and SPA methods, see Appendix \ref{appendice:generalized_SP_SPA}.

These advantages are retrieved using the proposed approach which draws a general framework to solve large-scale Bayesian inference problems.
Finally, as it will be shown in Section \ref{sec:Experiments}, the proposed SP and SPA algorithms need few fast iterations (akin to ADMM) to reach the same performance as state-of-the-art MCMC methods with good mixing properties. 

\section{Application to linear Gaussian inverse problems}
\label{sec:Bayesian_inference_problems}
In this section, the proposed splitting-and-augmenting strategy is envisioned to address two particular instances of linear Gaussian inverse problems formulated within a Bayesian framework. It first defines the considered class of problems and then derives the proposed approaches on two often-studied particular cases. Note that only the derivation of the SPA algorithm is discussed since it naturally embeds SP. However, the conclusions made hereafter stand also for SP. In Section \ref{sec:Experiments}, results of experiments associated to these two inverse problems will be reported and discussed.

\subsection{Linear Gaussian inverse problems}
Linear Gaussian inverse problems define an archetypal class of problems that could be efficiently tackled by the models and algorithms introduced in Sections \ref{sec:Model} and \ref{sec:Inference}. Suppose that some noisy signals $\B{y}$ are observed and one wants to infer an hidden parameter $\B{x}$ under the linear model
\begin{equation}\label{eq:LIP}
  \B{y} = \B{Hx} + \B{e}
\end{equation}
where $\B{H}$ is a direct operator and $\B{e}$ stands for noise or error modeling.
Then, assuming that $\B{e}$ is a Gaussian random vector with covariance matrix $\boldsymbol{\Omega}^{-1}$, the likelihood distribution associated with the observation vector $\B{y}$ is
\begin{equation}\label{eq:LIP_likelihood}
    p\pr{\B{y}|\B{x}} \propto \exp\br{-\dfrac{1}{2}\pr{\B{Hx}-\B{y}}^T\boldsymbol{\Omega}\pr{\B{Hx}-\B{y}}}.
\end{equation}
In most applicative contexts, $\B{H}$ is not invertible and inferring the unknown parameter vector $\B{x}$ from the observation vector $\B{y}$ under the linear model \eqref{eq:LIP} is known to be an ill-posed inverse problem. To alleviate this issue, a convenient and widely admitted approach consists in adopting some sort of regularization. Within a Bayesian setting, this is done by assigning a prior distribution to the unknown parameter vector $\B{x}$. Assuming that this prior distribution is given by the general form
\begin{equation}\label{eq:LIP_prior_x}
  p\pr{\B{x}} \propto \exp\br{-g(\B{x})},
\end{equation}
it follows by applying Bayes' rule that the posterior distribution of $\B{x}$ has the same form as (\ref{eq:targetDistribution}) where $f(\B{x}) = \dfrac{1}{2}\pr{\B{Hx}-\B{y}}^T\boldsymbol{\Omega}\pr{\B{Hx}-\B{y}}$.
As a consequence, the proposed methodology can be implemented to sample efficiently from a close approximation of this posterior distribution and use these samples to infer the hidden parameter $\B{x}$.
In the sequel, two standard problems involving Gaussian and total variation (TV) prior distributions, respectively, are considered.
One can easily verify that Assumptions \ref{assumption:1} and \ref{assumption:2} along with Theorem \ref{theorem:1} hold for all these problems.

\subsection{Deconvolution with a smooth prior}
\label{subsec:BayesProblems_Gaussian}
In the setup considered in this paragraph, the function $g$ in \eqref{eq:LIP_prior_x} is chosen to be quadratic as in \eqref{eq:g_quadratic} with $\boldsymbol{\mu}_2 = \B{0}_N$ and $\B{Q}_2 = \gamma\B{L}^T\B{L}$, where $\B{L}$ is a circulant matrix associated to a Laplacian filter. These choices lead to a frequently encountered smoothing conjugate Gaussian prior $\mathcal{N}\left({\B{0}_N, \pr{\gamma\B{L}^T\B{L}}}^{-1}\right)$, for instance used in \cite{Molina1989,Molina2006,Likas2004}. 
Note that this Gaussian prior distribution is degenerated since constant images are not penalized leading to the first eigenvalue of $\B{Q}_2$ being equal to zero.
Thus the posterior distribution (\ref{eq:gaussianPosterior}) becomes
\begin{equation}
  \label{eq:gaussianPosterior_bis}
  \pi(\B{x}|\B{y}) = \mathcal{N}\pr{\B{m},\B{Q}^{-1}}
\end{equation}
where
\begin{numcases}{}
\B{Q} = \B{H}^T\boldsymbol{\Omega}\B{H} + \gamma\B{L}^T\B{L} \label{eq:gaussianPosterior_cov}\\
\B{m} = \B{Q}^{-1}\B{H}^T\boldsymbol{\Omega}\B{y} \label{eq:gaussianPosterior_mean}.
\end{numcases}
Additionally, in the sequel, the operator $\B{H}$ will be assumed to be an $N \times N$ circulant convolution matrix associated to a time/space-invariant blurring kernel.
Finally, the noise covariance matrix is assumed to be diagonal, i.e., $\boldsymbol{\Omega}^{-1}=\mathrm{diag}[\sigma_1^2,\ldots,\sigma_N^2]$. Direct sampling according to the posterior distribution (\ref{eq:gaussianPosterior_bis}) is a challenging task, mainly due to the presence of the precision matrix $\boldsymbol{\Omega}$. Indeed, as emphasized in paragraph \ref{subsubsec:inference_gaussian}, the two terms in (\ref{eq:gaussianPosterior_cov}) cannot be diagonalized in the same basis (e.g. Fourier) which leads to computational problems in high dimension.

Conversely, assuming that $\phi_1$ and $\phi_2$ have the form (\ref{eq:phi1QuadraticBis}) and (\ref{eq:phi2Quadratic}) with  parameters $\rho$ and $\alpha$, the proposed SPA Gibbs algorithm samples according to the conditional distributions
  \begin{align}
    &p(\B{x}|\B{z},\B{u}) = \mathcal{N}\pr{\B{m_x},\B{G_x}^{-1}} \label{eq:gaussian_condDistrib1_x}\\
    &p(\B{z}|\B{x},\B{u}) = \mathcal{N}\pr{\B{m_z},\B{G_z}^{-1}} \label{eq:gaussian_condDistrib1_z}\\
    &p(\B{u}|\B{x},\B{z}) = \mathcal{N}\pr{\B{m_u},\B{G_u}^{-1}} \label{eq:gaussian_condDistrib_u}
  \end{align}
where
\begin{numcases}{}
\B{G_x} = \B{H}^T\boldsymbol{\Omega}\B{H} + \dfrac{1}{\rho^2}\B{I}_N \label{eq:gaussianPosterior_cov_x}\\
\B{G_z} =\gamma\B{L}^T\B{L} + \dfrac{1}{\rho^2}\B{I}_N \label{eq:gaussianPosterior_cov_z}\\
\B{G_u} = \dfrac{\alpha^2+\rho^2}{\alpha^2\rho^2}\B{I}_N.
\end{numcases}
Thanks to the splitting-and-augmenting approach, these three sampling steps are much easier to handle than the direct sampling from the target posterior distribution \eqref{eq:gaussianPosterior_bis}. Indeed, sampling from (\ref{eq:gaussian_condDistrib1_x}) can be conducted by using the auxiliary method of \cite{Marnissi2018} to deal separately with $\B{H^TH}$ from the coupling induced by $\boldsymbol{\Omega}$ (see Appendix \ref{appendice:efficientGaussianSampling_x}).
Additionally, sampling from (\ref{eq:gaussian_condDistrib1_z}) can be efficiently achieved in the Fourier domain (see Appendix \ref{appendice:efficientGaussianSampling_z} for details). Finally, sampling from \eqref{eq:gaussian_condDistrib_u} is straightforward since the covariance matrix is diagonal. Again, as previously noticed in Section \ref{sec:Inference} and more particularly in paragraph \ref{subsubsec:inference_gaussian} dedicated to Gaussian distributions, the proposed splitting-and-augmenting allows specific and simpler sampling
steps to be conducted where the difficulties inherent to $f$ (here the Gaussian likelihood) and $g$ (here the Gaussian prior) have been dissociated. The strategy developed in this paragraph will be experimentally assessed in paragraph \ref{subsec:Exp_Gaussian}.

\subsection{Image inpainting with total variation}
\label{subsec:BayesProblems_inpainting}
TV has become an ubiquitous regularization to solve imaging problems \cite{Rudin1992,Strong2003,Chambolle2010}. Within the considered Bayesian framework, it consists in choosing the $g$ function in \eqref{eq:LIP_prior_x} as $g(\B{x}) = \beta \mathrm{TV}(\B{x})$ where $\beta > 0$ and $\mathrm{TV}(\B{x}) = \sum_{1\leq i,j\leq N}\nr{\pr{\nabla\B{x}}_{i,j}}_2$ ($\nabla\B{x}$ is the two-dimensional discrete gradient of $\B{x}$). This type of prior is used for instance in image inpainting problems, which consist in recovering an original image $\B{x} \in \mathbb{R}^N$ from the noisy and partial measurements $\B{y} \in \mathbb{R}^M$ under the linear model (\ref{eq:LIP}). Note that, in general, $M\ll N$.
Here, the noise is assumed to be white and Gaussian such that $\boldsymbol{\Omega}^{-1} = \sigma^2\B{I}_M$ and the operator $\B{H}$ stands for the matrix associated with a damaging binary mask. Under this setting, the posterior distribution of $\B{x}$ (\ref{eq:targetDistribution}) becomes
\begin{equation}\label{eq:inpainting_likelihood}
  p\pr{\B{x}|\B{y}} \propto \exp\br{-\dfrac{1}{2\sigma^2}\nr{\B{Hx}-\B{y}}^2_2 - \beta \mathrm{TV}(\B{x})}.
\end{equation}
Direct sampling from this posterior is a challenging task mainly due to \emph{i)} the generally high dimension of the image to be recovered, \emph{ii)} the non-conjugacy of the TV-based prior, leading to a non-standard posterior distribution and \emph{iii)} the non-differentiability of $g$ which precludes the use of some advanced simulation techniques, e.g., Hamiltonian Monte Carlo algorithms \cite{Duane1987}. Conversely, instead of directly sampling from this posterior distribution, the proposed approach is applied.
Again, assuming that $\phi_1$ and $\phi_2$ have the forms (\ref{eq:phi1QuadraticBis}) and (\ref{eq:phi2Quadratic}) with parameters $\rho$ and $\alpha$, respectively, the conditional distributions associated to SPA are
  \setlength{\arraycolsep}{0.0em}%
  \begin{eqnarray}
    p(\B{x}|\B{z},\B{u})&{} \propto&{} \exp\br{-\dfrac{1}{2\sigma^2}\nr{\B{Hx}-\B{y}}^2_2} \nonumber \\
    &&\times \exp\br{- \dfrac{1}{2\rho^2}\nr{\B{x}-(\B{z}-\B{u})}^2_2} \label{eq:inpainting_condDistrib_x}\\
    p(\B{z}|\B{x},\B{u})&{} \propto&{} \exp\br{-\beta \mathrm{TV}(\B{z}) - \dfrac{1}{2\rho^2}\nr{\B{z}-(\B{x}+\B{u})}^2_2} \label{eq:inpainting_condDistrib_z}\\
    p(\B{u}|\B{x},\B{z})&{} \propto&{} \exp\br{-\dfrac{1}{2\alpha^2}\nr{\B{u}}^2_2 - \dfrac{1}{2\rho^2}\nr{\B{u}-(\B{z}-\B{x})}^2_2} \label{eq:inpainting_condDistrib_u}
  \end{eqnarray}
\setlength{\arraycolsep}{5pt}%
Here, assuming that $\phi_1$ and $\phi_2$ are quadratic allows to retrieve Gaussian distributions for (\ref{eq:inpainting_condDistrib_x}) and (\ref{eq:inpainting_condDistrib_u}).
Sampling from (\ref{eq:inpainting_condDistrib_u}) in high-dimension is not a problem since the covariance matrix is constant diagonal.
However, the covariance matrix associated to (\ref{eq:inpainting_condDistrib_x}) is $\pr{\sigma^{-2}\B{H}^T\B{H} + \rho^{-2}\B{I}_N}^{-1}$, which is more complex to handle.
Hopefully, the direct operator $\B{H}$ is a $M \times N$ binary matrix which can be obtained by taking a subset of rows of the identity matrix in dimension $N$.
Due to this simple structure, $\B{H}\B{H}^T = \B{I}_M$ and by using the Sherman-Morrison-Woodbury formula, it follows that
\begin{align}
  \pr{\dfrac{1}{\sigma^2}\B{H}^T\B{H} + \dfrac{1}{\rho^2}\B{I}_N}^{-1} = \rho^2\pr{\B{I}_N - \dfrac{\rho^2}{\sigma^2 + \rho^2}\B{H}^T\B{H}}. \label{eq:cov_inpainting}
\end{align}
The matrix $\B{H}^T\B{H}$ corresponds to an identity matrix with some zeros in the diagonal (corresponding to the missing pixels). Thereby, the covariance matrix (\ref{eq:cov_inpainting}) is diagonal and the sampling from (\ref{eq:inpainting_condDistrib_x}) can be conducted efficiently with the exact perturbation-optimization (E-PO) algorithm \cite{Papandreou2011}.

As previously discussed in paragraph \ref{subsubsec:inference_logconcave}, the conditional distribution (\ref{eq:inpainting_condDistrib_z}) being log-concave, one can sample efficiently from the latter in high-dimension with P-MALA or P-MYULA. In the sequel, P-MYULA will be preferred because its mixing properties are better than P-MALA and the estimation error is of the order of 1\% using well-defined parameters \cite{Durmus2018}. As a conclusion, as advocated earlier, the proposed splitting-and-augmenting approach allows simpler sampling steps to be efficiently conducted thanks to dedicated algorithms. 

\section{Experiments}
\label{sec:Experiments}
This section reports results of experiments aimed at comparing the proposed methodology with that of current state-of-the-art (optimization and Bayesian) methods for the inverse problems discussed in Section \ref{sec:Bayesian_inference_problems}. All the results presented in this section have been obtained using MATLAB, on a computer equipped with an Intel Xeon 3.70 GHz processor, with 16.0 GB of RAM, and running Windows 7.
Other examples of the proposed approach on machine learning problems can be found in \cite{Rendell2018,Vono_MLSP18}.

\subsection{Deconvolution with a smooth prior}
\label{subsec:Exp_Gaussian}
\subsubsection{Problem considered}
The Gaussian sampling problem introduced in Section \ref{subsec:BayesProblems_Gaussian} is considered.
A blurred and noisy image $\B{y} \in \mathbb{R}^M$ of size $512 \times 512$ ($M = 262 144$) is observed. The purpose is then to recover the original image $\B{x} \in \mathbb{R}^N$ of size $512 \times 512$ ($N = 262 144$).

\subsubsection{Experimental design}
The diagonal elements $\sigma_i^2$ of the noise covariance matrix $\boldsymbol{\Omega}^{-1}$ have been randomly drawn according to the mixture $\sigma_i \sim (1-\beta)\delta_{\kappa_1} + \beta\delta_{\kappa_2}$ ($\kappa_1, \kappa_2 > 0$ and $0< \beta <1$) with $\beta = 0.35$, $\kappa_1 = 13$ and $\kappa_2 = 40$.
This particular structure for $\boldsymbol{\Omega}^{-1}$ may be not physical but permits to show the interest of the proposed approach.
The prior parameter $\gamma$ has been set to $\gamma = 6 \times 10^{-3}$.

The proposed SP and SPA algorithms SP are compared to RJ-PO \cite{Gilavert2015} and to the algorithms denoted AuxV1 and AuxV2 proposed in \cite{Marnissi2018}. The parameters associated to SP and SPA have been set to $\rho = 20$ and $(\rho, \alpha) = (20,1)$, respectively. RJ-PO has been run using conjugate gradient (CG) algorithm as the required linear solver whose tolerance has been adapted to reach an acceptance rate of $0.9$. The number of burn-in iterations has been set to $T_{\mathrm{bi}} = 200$ for AuxV1, RJ-PO, SP and SPA and to $T_{\mathrm{bi}} = 2200$ for AuxV2 (due to its slower mixing properties, see below). For each MCMC algorithm, $800$ samples obtained after the burn-in period have been used.
The number of iterations $T_{\mathrm{MC}}$ and $T_{\mathrm{bi}}$ were empirically chosen by graphically inspecting the behavior of the Markov chains produced by the samplers.

The performances of the different approaches have been assessed by the signal-to-noise ratio (SNR) and the peak signal-to-noise ratio (PSNR)
\begin{align}
  &\mathrm{SNR} = 10 \log_{10} \dfrac{\nr{\B{x}}_2^2}{\nr{\B{x} - \hat{\B{x}}}_2^2} \\
  &\mathrm{PSNR} = 10 \log_{10} \dfrac{255^2}{N^{-1}\nr{\B{x} - \hat{\B{x}}}_2^2}
\end{align}
where $\hat{\B{x}}$ refers to the MMSE estimate of $\B{x}$ approximated by empirical averages of the samples generated by the MCMC algorithms. The performance results have been averaged over $25$ Monte Carlo runs.

\subsubsection{Results}
Table \ref{table:gaussian_MSE} shows the average SNR and PSNR associated to the MMSE estimate for the different algorithms. The standard deviation associated to these results is the same for the different methods and is equal to $0.02$ and all the algorithms share similar performance results. However, we emphasize that the computational cost of each algorithm can differ widely as shown by Table \ref{table:gaussian}.

\begin{table}
\renewcommand{\arraystretch}{1.3}
\caption{Gaussian sampling: average SNR and PSNR (over 25 observations) associated to the MMSE estimates.}
\label{table:gaussian_MSE}
\centering
\begin{tabular}{l l l}
\thickhline
\bfseries & \bfseries SNR (dB) & \bfseries PSNR (dB)\\
\hline
RJ-PO & 19.58 & 25.24 \\
AuxV1 & 19.58 & 25.24 \\
AuxV2 & 19.60 & 25.26 \\
SP & 19.58 & 25.23 \\
SPA & 19.58 & 25.23 \\
\thickhline
\end{tabular}
\end{table}

\begin{figure}
\centering
\includegraphics[width=3.5in]{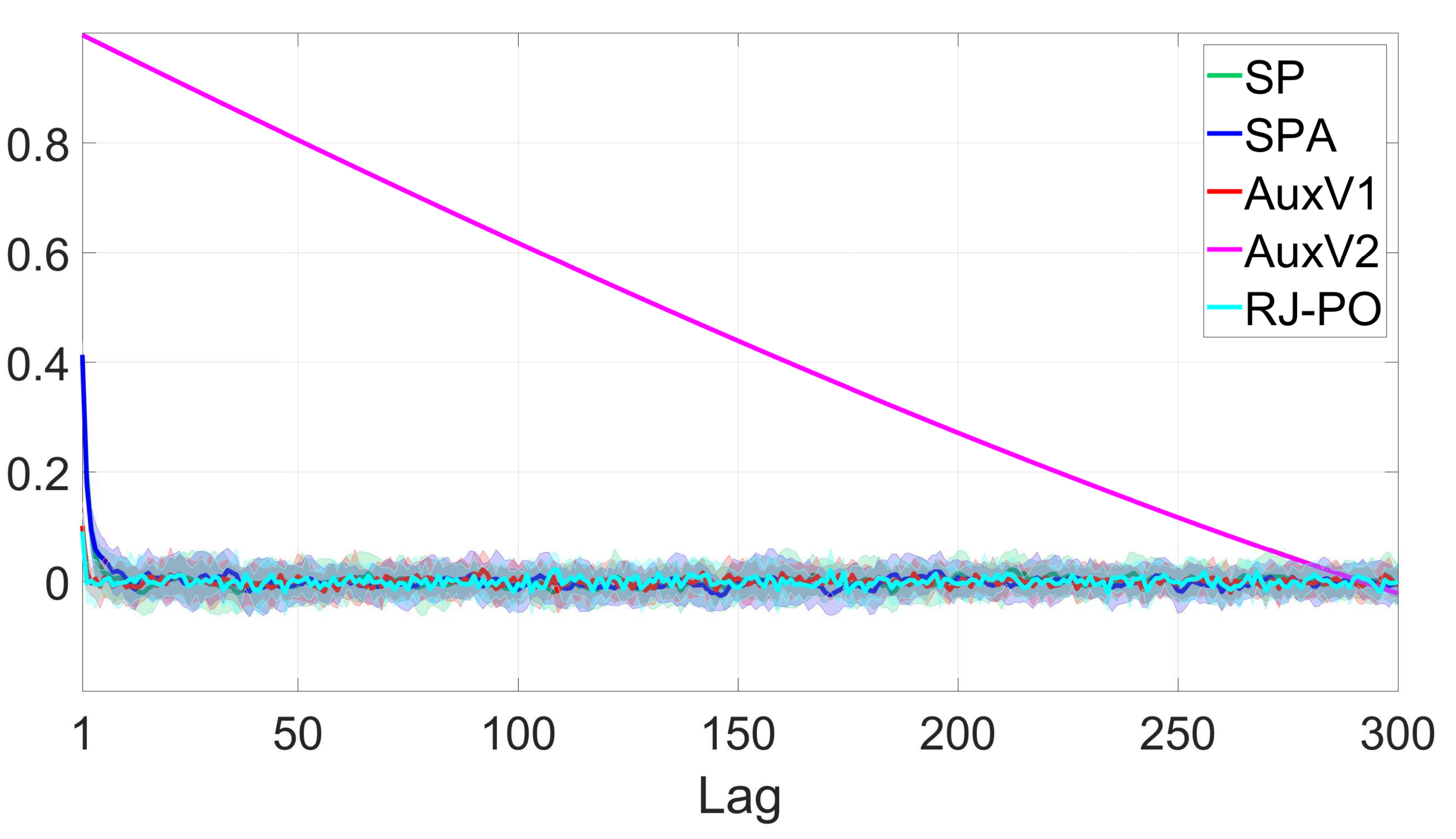}
\caption{Gaussian sampling: average chain autocorrelation functions of SP (green), SPA (blue), AuxV1 (red), AuxV2 (magenta) and RJ-PO (cyan).
Shaded areas represent the intervals corresponding to the standard deviation computed over 25 trials.}
\label{fig:gaussian_ACF}
\end{figure}

Table \ref{table:gaussian} presents the numerical complexity related to one iteration of each algorithm along with the average number of iterations performed and the average computational time for each algorithm (over the $25$ Monte Carlo runs). The complexity of $\mathcal{N}$ refers to the sampling from an univariate normal distribution. The complexity of $O(N\log N)$ refers to the use of the Fourier transform as the matrices $\B{H}$ and $\B{L}$ are circulant and thereby diagonalizable in the Fourier domain.
One can denote that SP, SPA, AuxV1 and AuxV2 share a roughly similar numerical complexity (for one iteration) whereas RJ-PO is slower because of the use of the CG method.
The latter has a complexity of $O(N_{\mathrm{CG}}N\log N)$ where $N_{\mathrm{CG}}$ is the number of iterations performed by the CG method.
In this example, $N_{\mathrm{CG}} = 155$ on average (after the burn-in period).
On the other hand, the average computing times associated to each MCMC algorithm widely differ.
RJ-PO is the slowest mainly due to the number of CG iterations performed at each iteration.
AuxV1 appears to be the fastest. However, one has to recall that this algorithm was explicitly designed for this type of inference problems and cannot be used directly for more general Gaussian sampling tasks.
SP and SPA appear to have reasonable computational costs compared to AuxV1.
Finally, AuxV2 needs more iterations and thereby more time to reach the same level of performance as the other approaches.
This algorithm can be used in more general cases than AuxV1 but appears to be roughly 3 times more costly than the proposed approach which covers a wider scope of sampling problems.
This high computational cost is mainly related to the poor mixing properties of AuxV2 compared to the other methods as drawn by Fig. \ref{fig:gaussian_ACF}.

\begin{table}
\renewcommand{\arraystretch}{1.3}
\caption{Gaussian sampling: computational complexity related to one iteration, average number of iterations and average computational time for each algorithm.}
\label{table:gaussian}
\centering
\begin{tabular}{l l l l}
\thickhline
\bfseries & \bfseries computational complexity & \bfseries $\#$ iterations & \bfseries time (s) \\
\hline
RJ-PO & $O(N_{\mathrm{CG}}N\log N) + (M+N)\mathcal{N}$ & $10^3$ & 4192 \\
AuxV1 & $O(N\log N) + 2N\mathcal{N}$ & $10^3$ & 37\\
AuxV2 & $O(N\log N) + 4N\mathcal{N}$ & $3 \times 10^3$ & 209\\
SP & $O(N\log N) + 3N\mathcal{N}$ & $10^3$ & 62\\
SPA & $O(N\log N) + 4N\mathcal{N}$ & $10^3$ & 86\\
\thickhline
\end{tabular}
\end{table}

Fig. \ref{fig:gaussian_ACF} compares the autocorrelation functions (using $-\log \pi(\B{x}|\B{y})$ as a scalar summary) of AuxV1, AuxV2, RJ-PO, SP and SPA averaged over the $25$ Monte Carlo runs, where only samples obtained after the burn-in period have been considered.
The shaded regions depicted in Fig. \ref{fig:gaussian_ACF} represent the standard deviation ranges associated to each MCMC algorithm.
One can denote that all the algorithms share good mixing properties except AuxV2 which explores less efficiently the parameter space.
This result is consistent with the findings highlighted in \cite{Marnissi2018} which  pointed out that the quality of the samples generated by RJ-PO and AuxV1 was better than those generated by AuxV2.

\subsubsection{Discussion}
For this specific experiment, the proposed general splitting-and-augmenting framework has shown that it can compete with efficient algorithms designed only for this type of sampling problems (e.g. AuxV1).
Additionally, it proves to be more efficient than algorithms designed for wider Gaussian sampling tasks (e.g. AuxV2 and RJ-PO).
The performance of the proposed approach is strengthened by the fact that SP and SPA have also demonstrated to be more efficient than state-of-the-art MCMC algorithms designed to sample from other types of distributions, such as log-concave densities, as illustrated in the next paragraph \ref{subsec:Exp_inpainting}.

\subsection{Image inpainting with total variation}
\label{subsec:Exp_inpainting}
\subsubsection{Problem considered}
\begin{figure}
\centering
  \mbox{{\includegraphics[height=1.5cm]{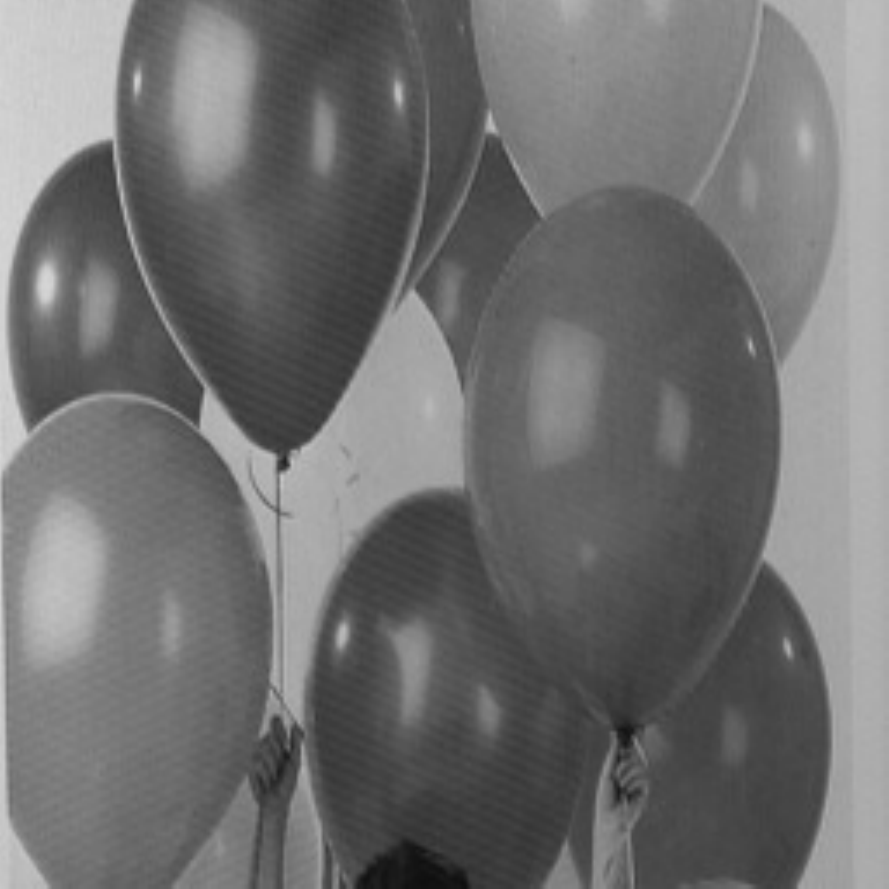}}}
  \mbox{{\includegraphics[height=1.5cm]{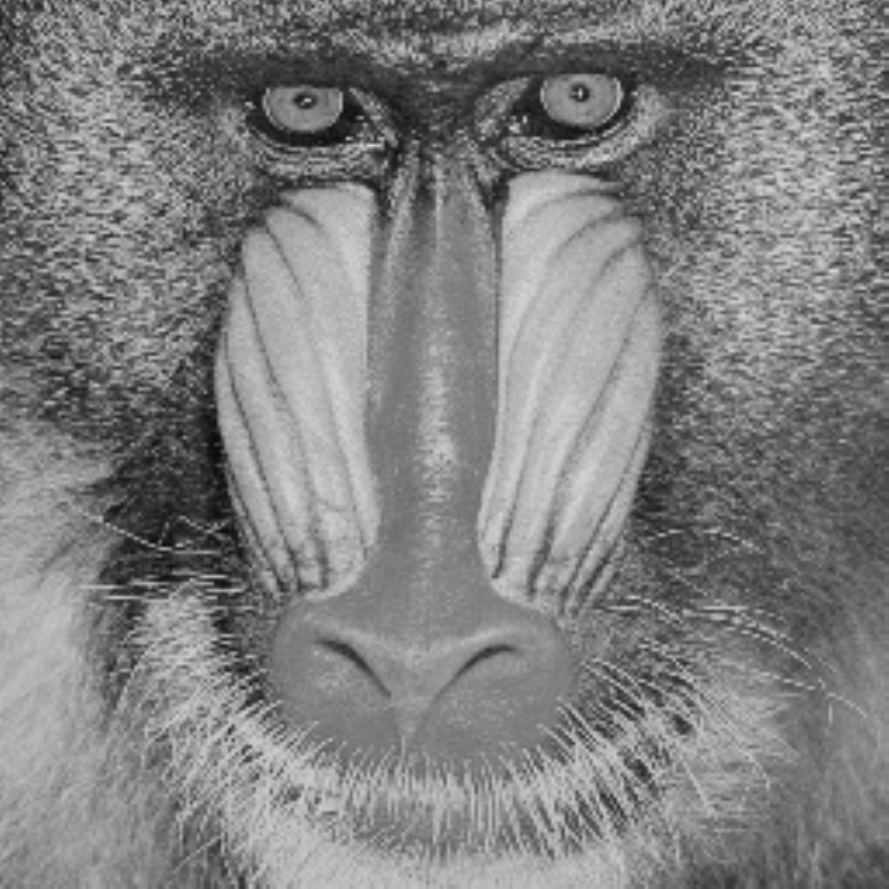}}}
  \mbox{{\includegraphics[height=1.5cm]{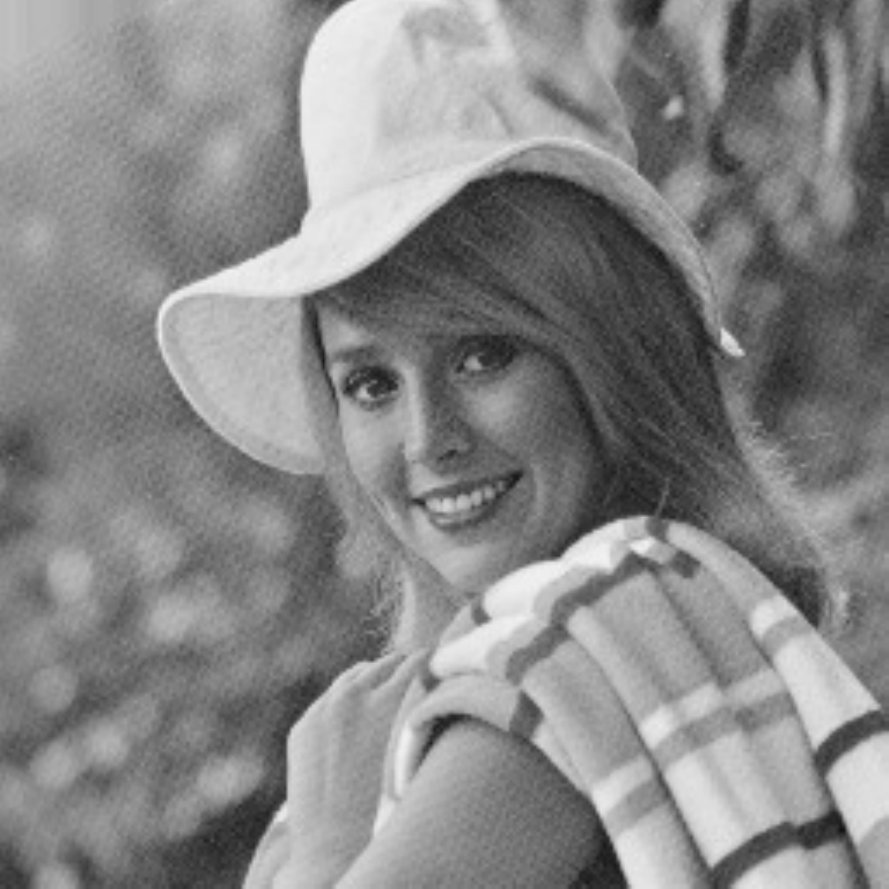}}}
  \mbox{{\includegraphics[height=1.5cm]{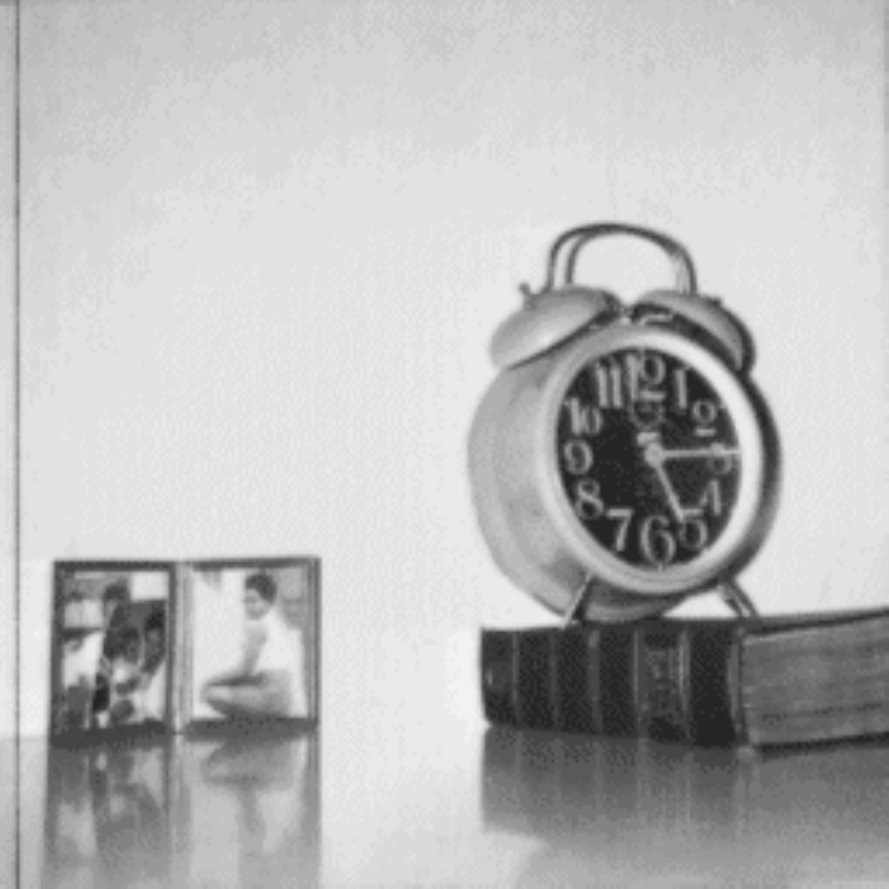}}}
  \mbox{{\includegraphics[height=1.5cm]{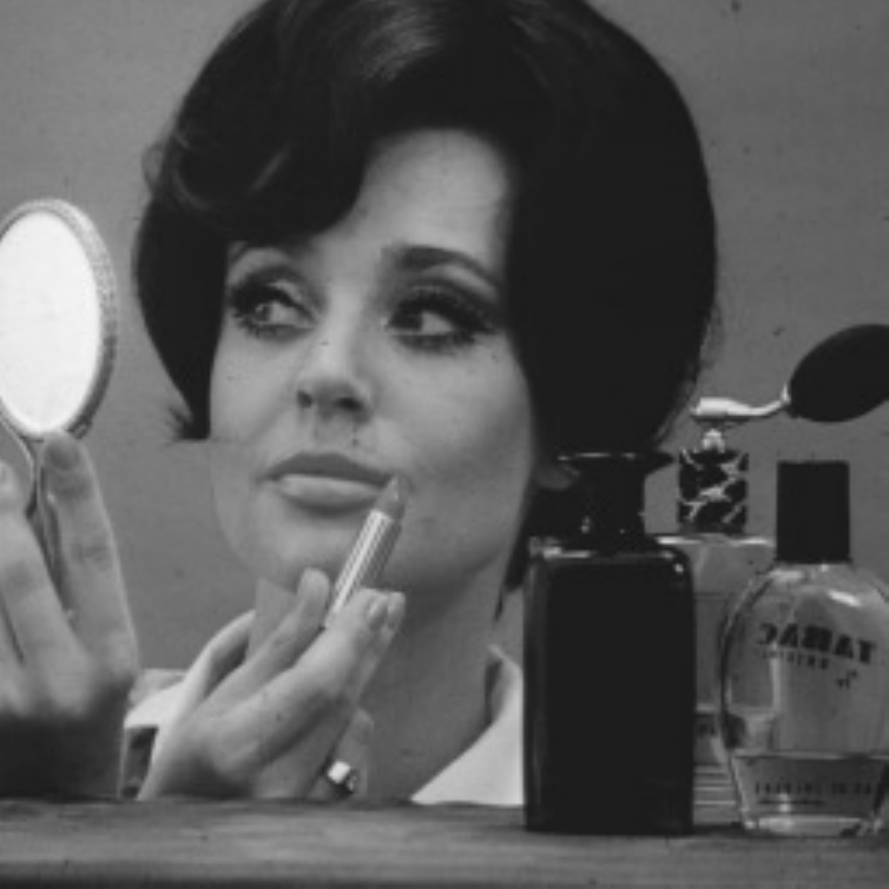}}}\vspace{0.1cm}
  \mbox{{\includegraphics[height=1.5cm]{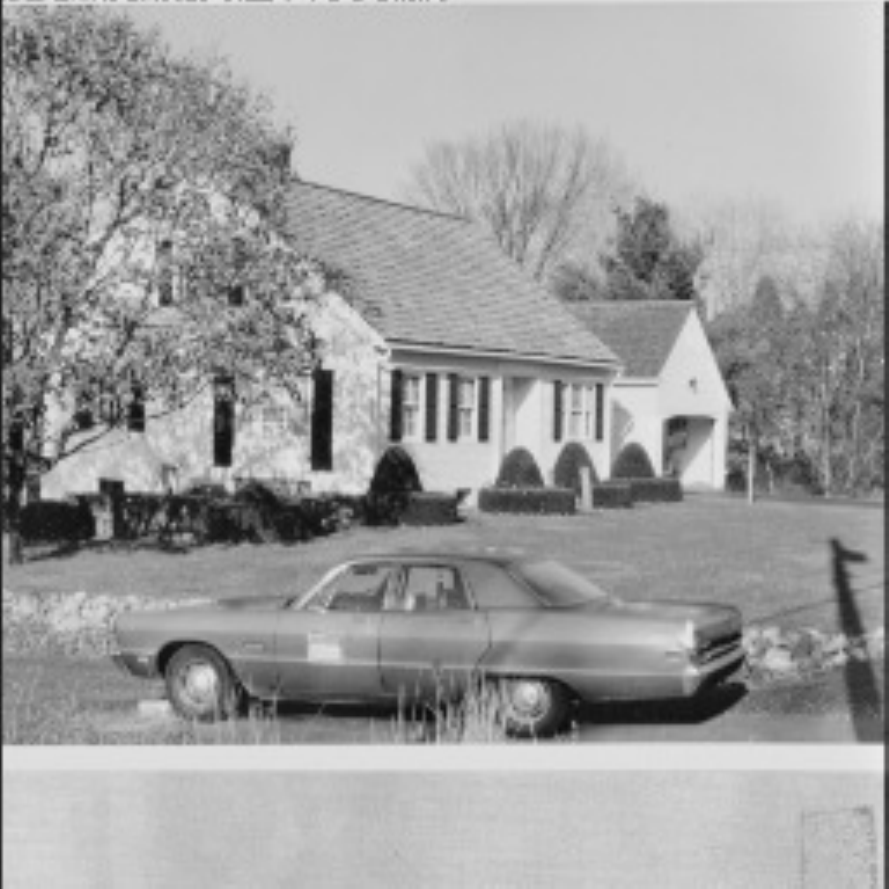}}}
  \mbox{{\includegraphics[height=1.5cm]{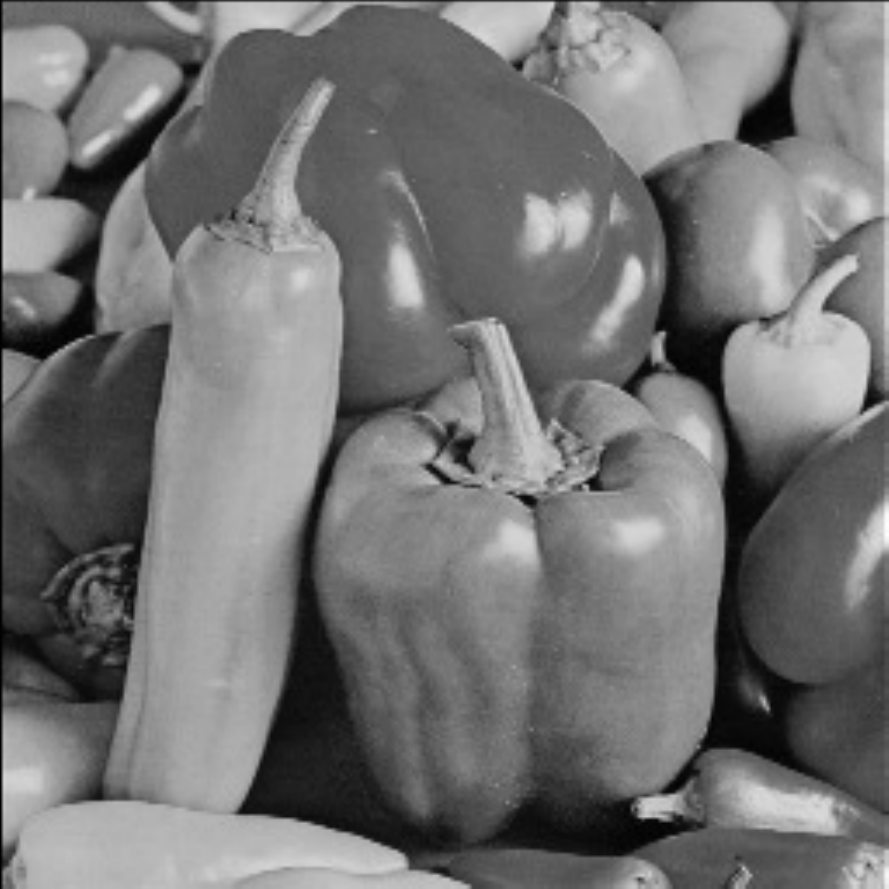}}}
  \mbox{{\includegraphics[height=1.5cm]{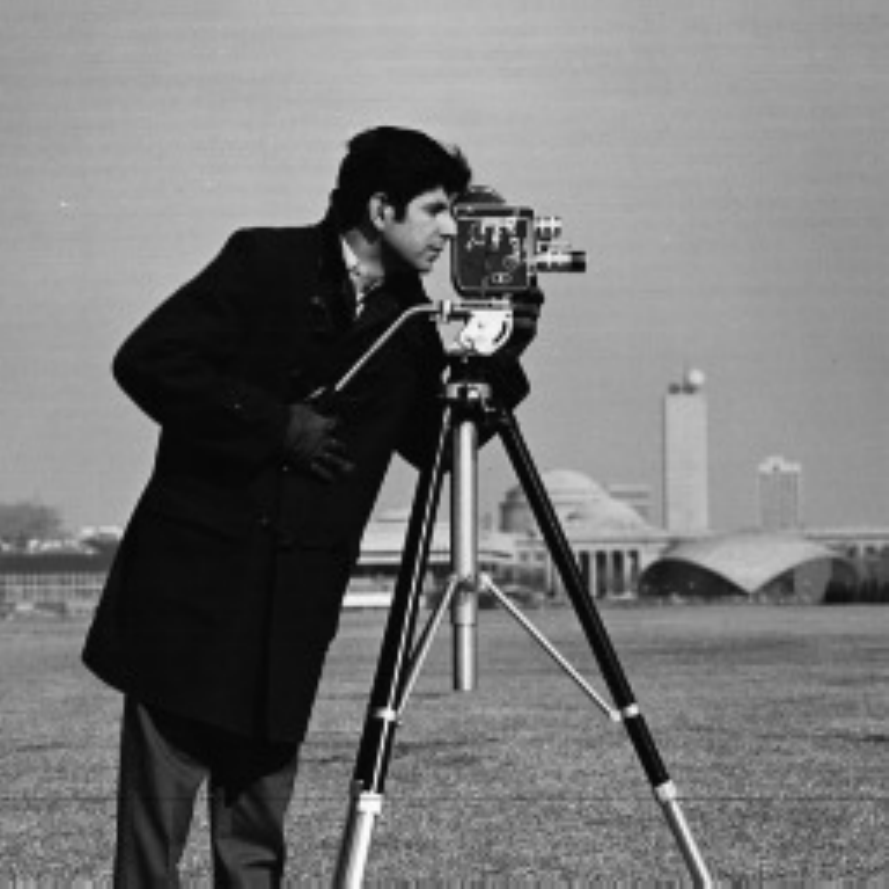}}}
  \mbox{{\includegraphics[height=1.5cm]{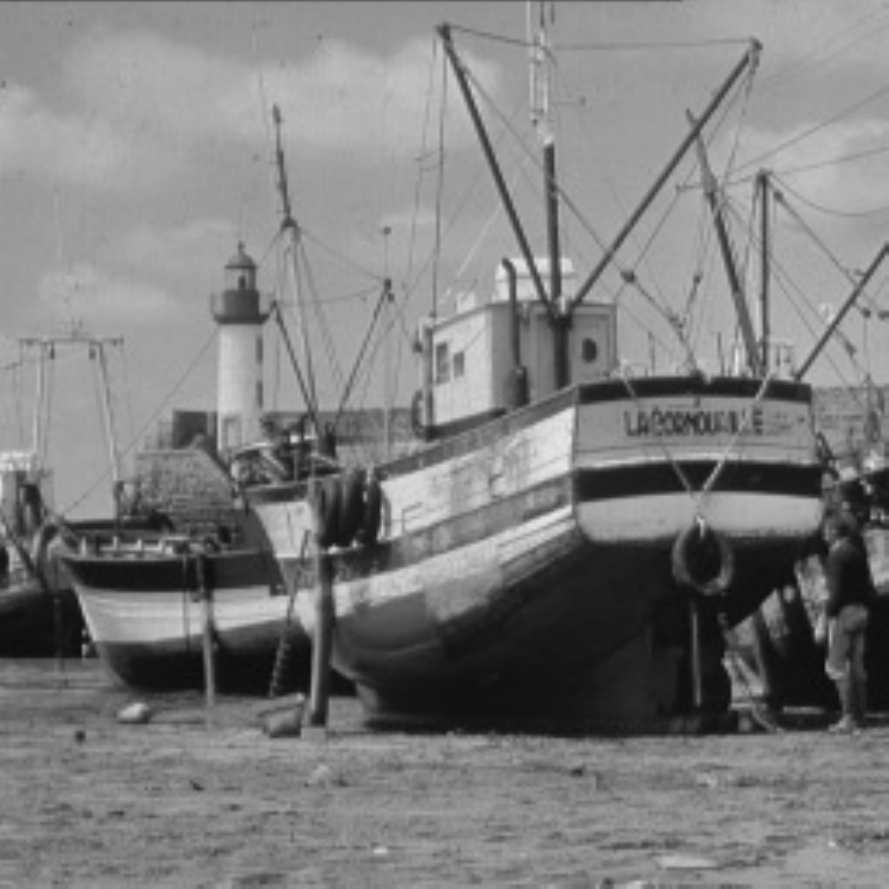}}}
  \caption{Set of $256 \times 256$ images used. From top left to bottom right: balloons, baboon, elaine, clock, donna, house, peppers, cameraman, boat.}
  \label{fig:9_images}
\end{figure}

The image inpainting problem introduced in Section \ref{subsec:BayesProblems_inpainting} and also addressed in \cite{Afonso2010} is considered here. Fig. \ref{fig:9_images} presents the nine $256 \times 256$ original gray-level images used for this experiment. The observation vector denoted $\B{y}$ consists of $60$\% randomly selected of the original image pixels $\B{x}$, corrupted by a white Gaussian noise with SNR of $40$dB.

\begin{figure}
\centering
  \mbox{\subfloat[]{\label{subfig:cameraman_x} \includegraphics[height=4.2cm]{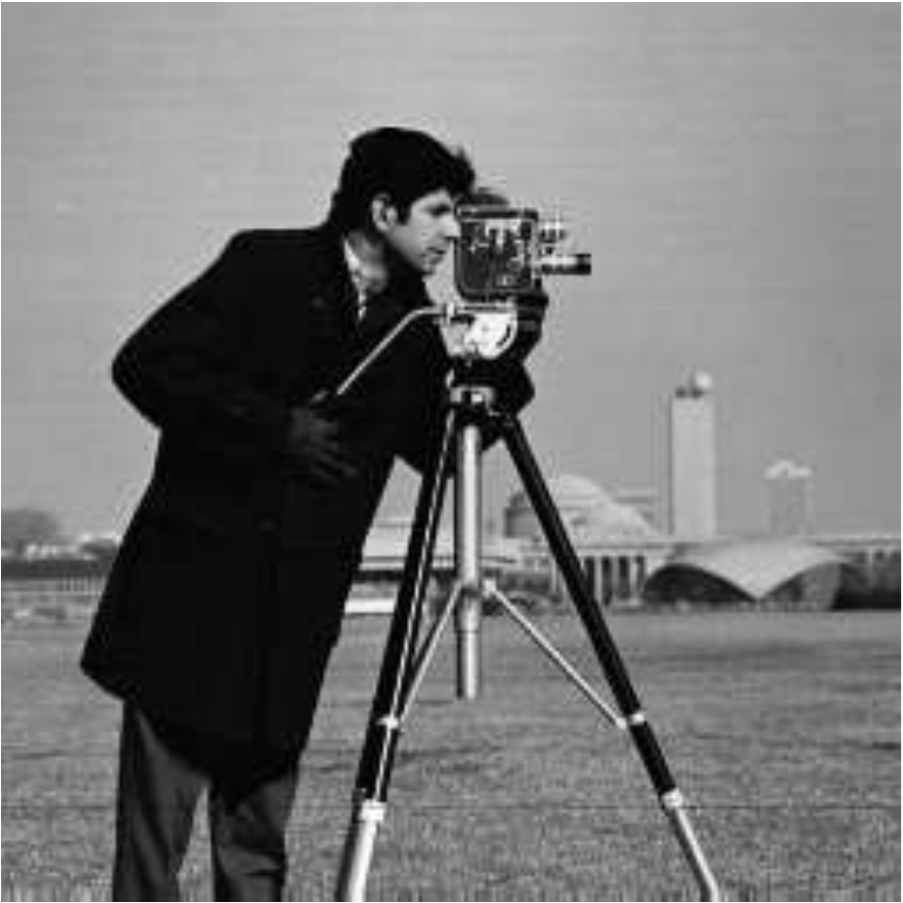}}}
  \mbox{\subfloat[]{\label{subfig:cameraman_y} \includegraphics[height=4.2cm]{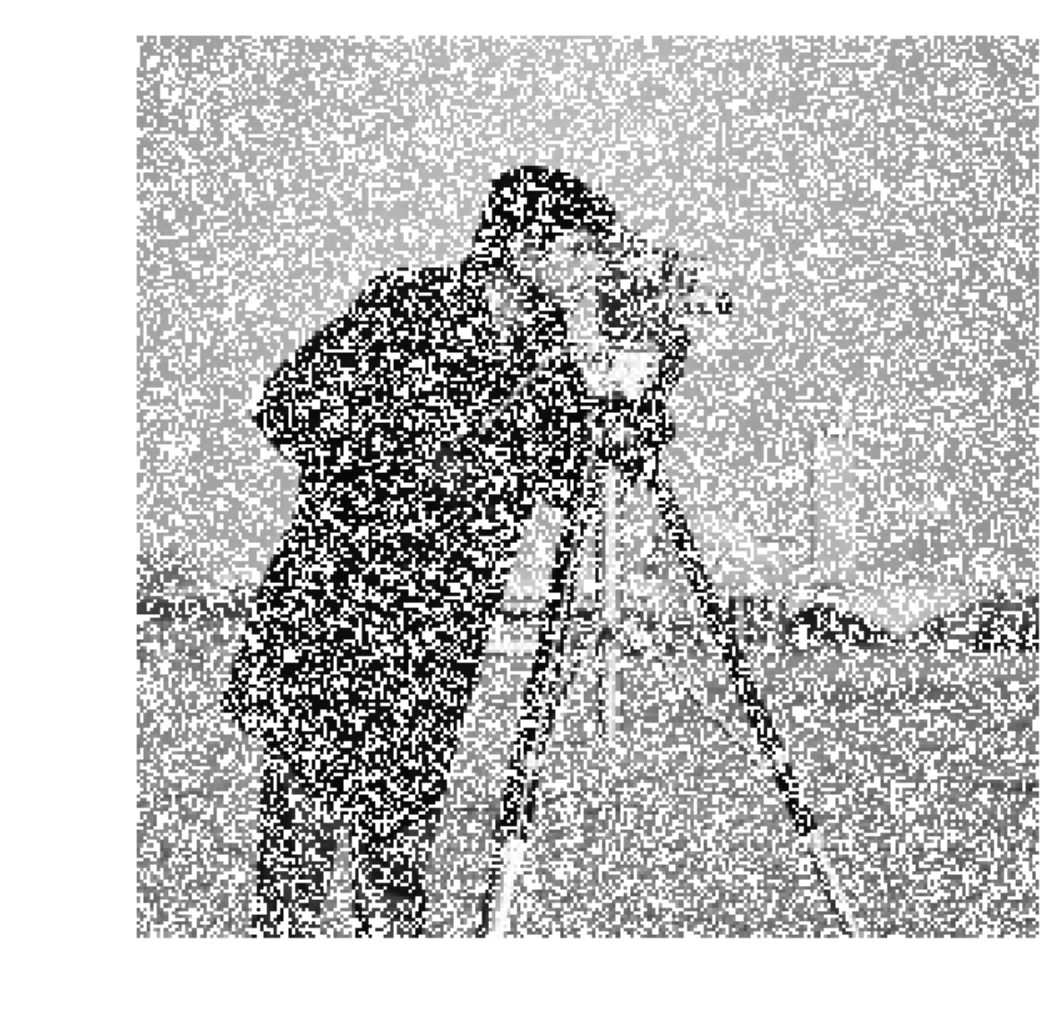}}}
  \mbox{\subfloat[]{\label{subfig:cameraman_SPA_MMSE_x} \includegraphics[height=4.2cm]{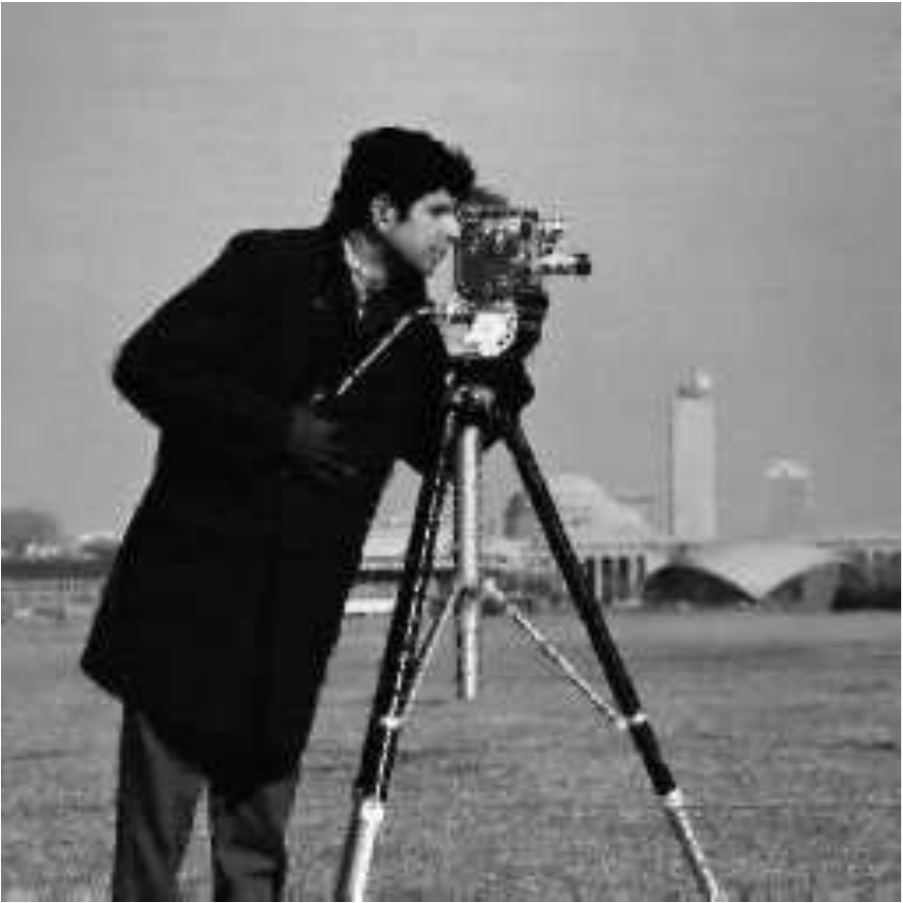}}}
  \mbox{\subfloat[]{\label{subfig:cameraman_SPA_MMSE_z} \includegraphics[height=4.2cm]{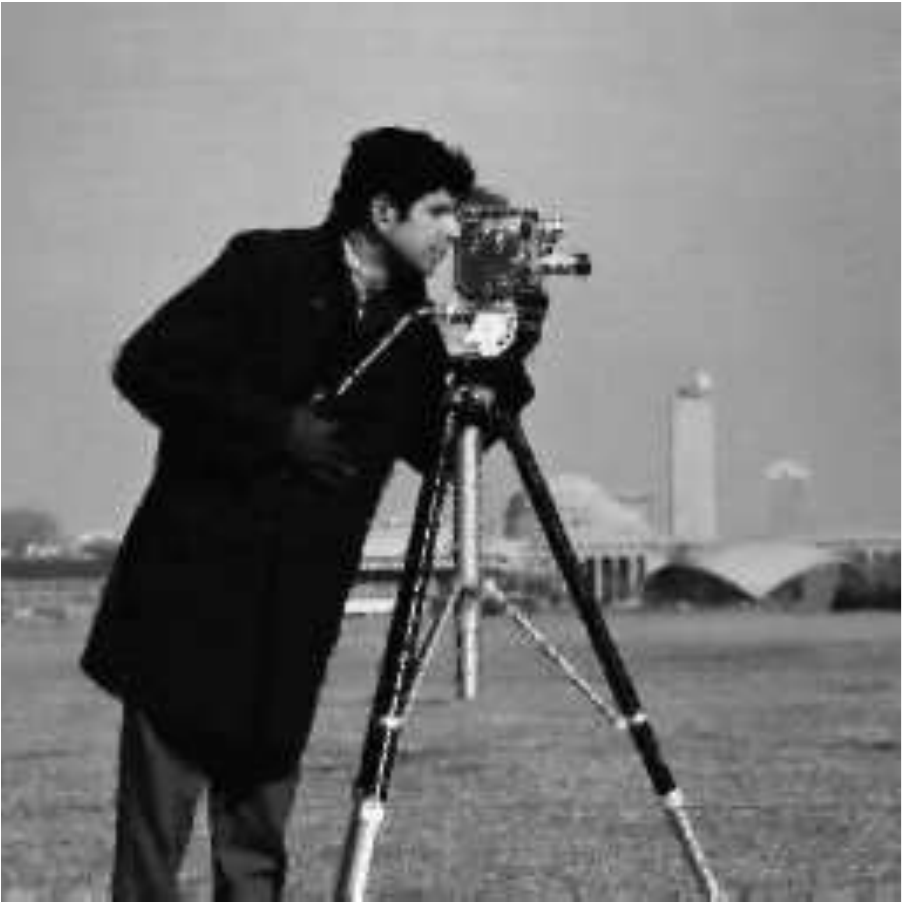}}}
  \mbox{\subfloat[]{\label{subfig:cameraman_SPA_MMSE_z} \includegraphics[height=3.4cm]{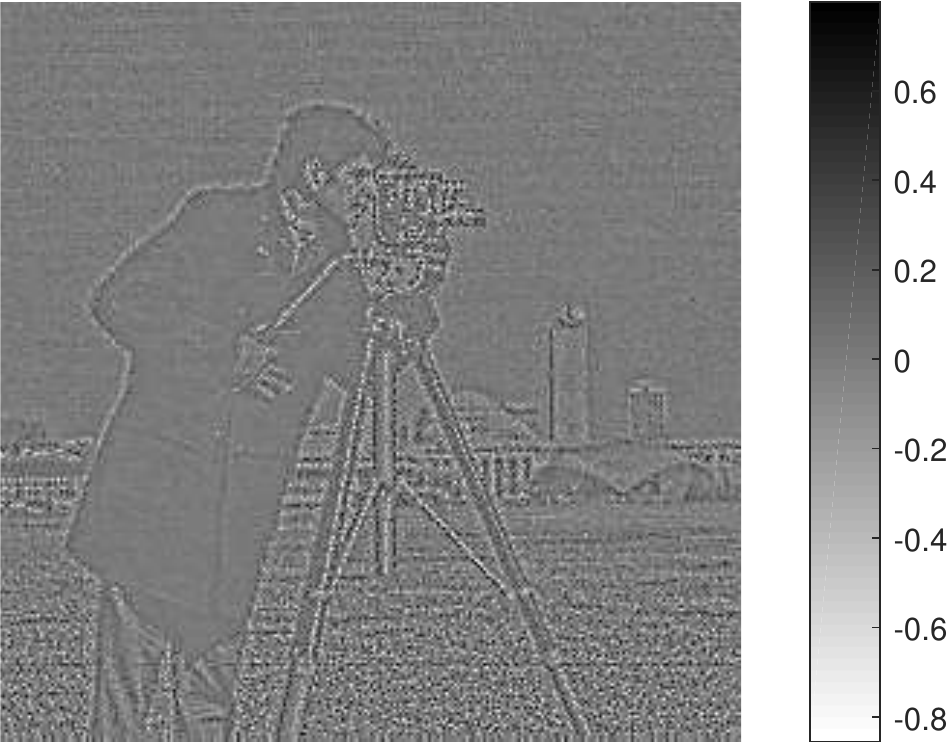}}}
  \mbox{\subfloat[]{\label{subfig:cameraman_SPA_CI90} \includegraphics[height=3.4cm]{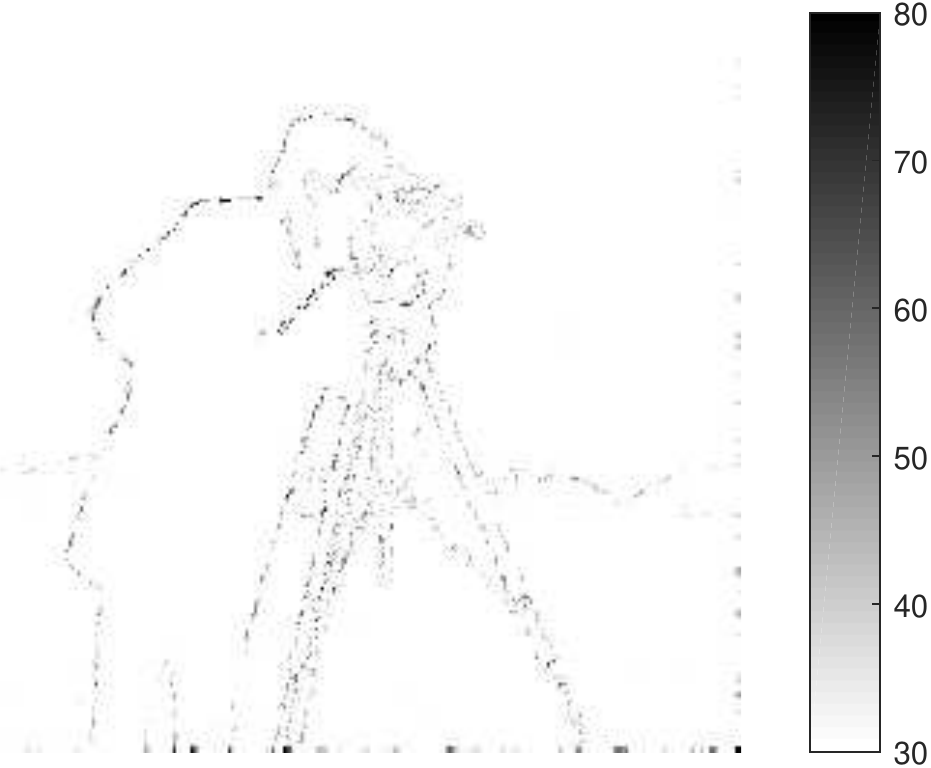}}}
  \caption{Image inpainting with TV regularization using SPA: (a) original image; (b) noisy observation with missing pixels depicted in white; (c) MMSE estimate of $\B{x}$; (d) MMSE estimate of $\B{z}$; (e) MMSE estimate of $\B{u}$; (f) Pixel-wise 90\% credibility intervals.}
  \label{fig:cameraman}
\end{figure}

Fig. \ref{subfig:cameraman_x} and \ref{subfig:cameraman_y} present, as an example, the original Cameraman image and one of its associated observations where the missing pixels are depicted in white. The restoration results for this image are also presented in Fig. \ref{fig:cameraman}.

\subsubsection{Experimental design}
The two proposed algorithms SP and SPA, leading to sampling from (\ref{eq:inpainting_condDistrib_x})-(\ref{eq:inpainting_condDistrib_u}), are compared with the split augmented Lagrangian shrinkage algorithm (SALSA) \cite{Afonso2010}, which can be interpreted as a deterministic counterpart of SPA, as emphasized in paragraph \ref{subsec:SPA_meets_ADMM}. SALSA solves the minimization problem resulting from the MAP inference associated with the posterior distribution \eqref{eq:inpainting_likelihood} by using ADMM. These algorithms have been also compared with P-MYULA specifically designed to sample from possibly non-smooth log-concave distributions (see paragraph \ref{subsubsec:inference_logconcave}). The number of burn-in iterations has been set to $T_{\mathrm{bi}} = 200$ for SP and SPA and to $T_{\mathrm{bi}} = 95200$ for P-MYULA (due to slower mixing, see below). For each MCMC algorithm, $4800$ samples obtained after the burn-in period have been used to approximate the MMSE estimator by empirical averaging.

Sampling from (\ref{eq:inpainting_condDistrib_z}) has been done with P-MYULA ($\lambda = \rho^2$ and $\gamma = \rho^2/4$) using Chambolle's algorithm \cite{Chambolle2004} to compute the proximal operator of $g$. The SP and SPA parameters have been set to $\rho = 2.8$, $\alpha = 1$ and $\beta = 0.2$ for Algo. \ref{algo:Gibbs1} and to $\rho = 2$ and $\beta = 0.2$ for Algo. \ref{algo:Gibbs2}. In particular, the choice of $\rho$ is discussed thereafter.

The performance of the estimators has been measured by computing the improvement in signal-to-noise ratio (ISNR) defined as
\setlength{\arraycolsep}{0.0em}
  \begin{align}
    &\mathrm{ISNR} = 10\log_{10}\dfrac{\nr{\B{x}-\B{y}}_2^2}{\nr{\B{x}-\hat{\B{x}}}_2^2}
  \end{align}
\setlength{\arraycolsep}{5pt}
where $\hat{\B{x}}$ refers to the MMSE (resp. MAP) estimate of $\B{x}$ for SP, SPA and P-MYULA (resp., SALSA). This performance measure has been averaged over $25$ Monte Carlo runs.

\subsubsection{Influence of $\alpha$}
\begin{figure}
  \centering
  \mbox{{\includegraphics[scale=0.55]{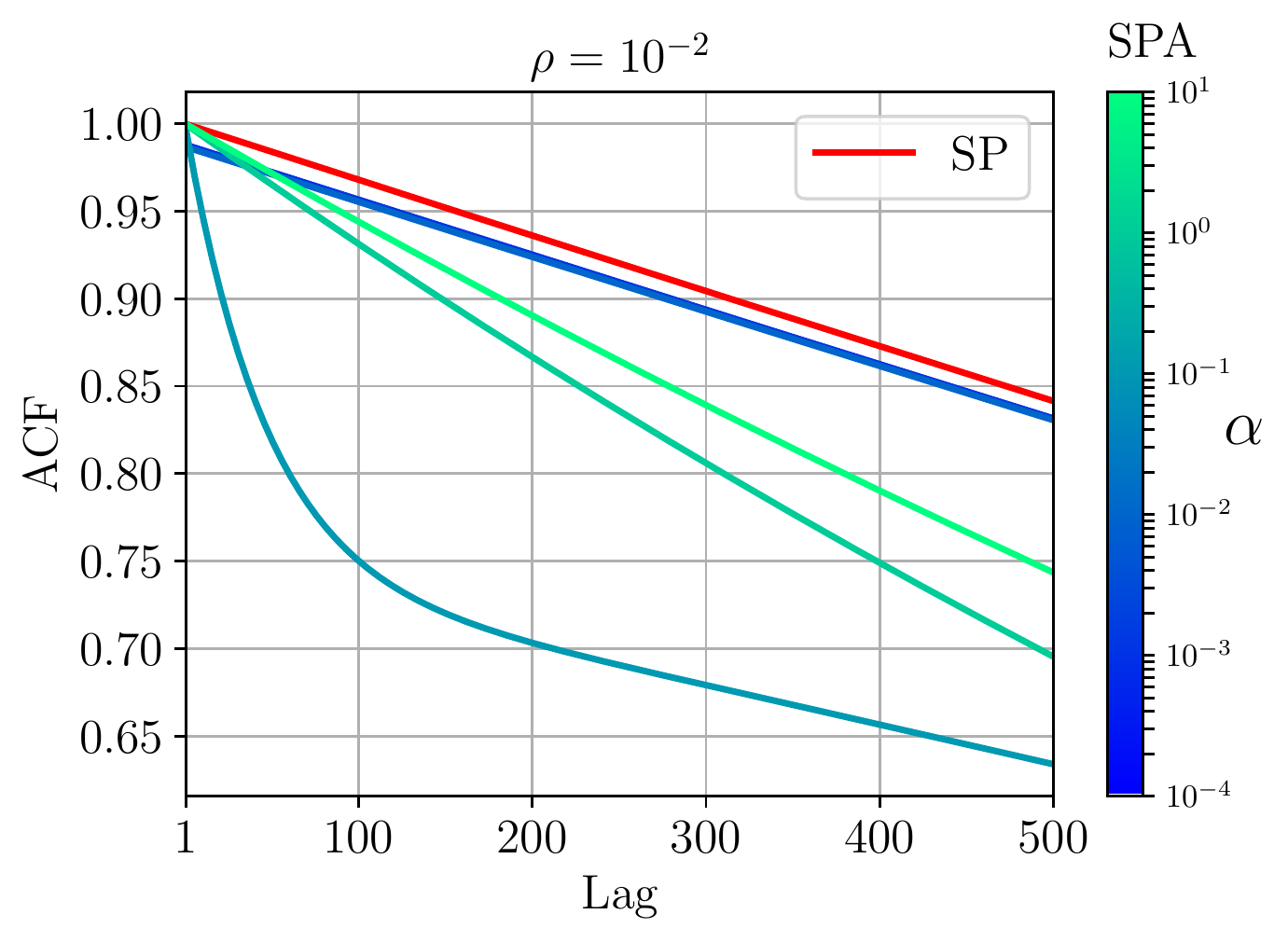}}}
  \mbox{{\includegraphics[scale=0.55]{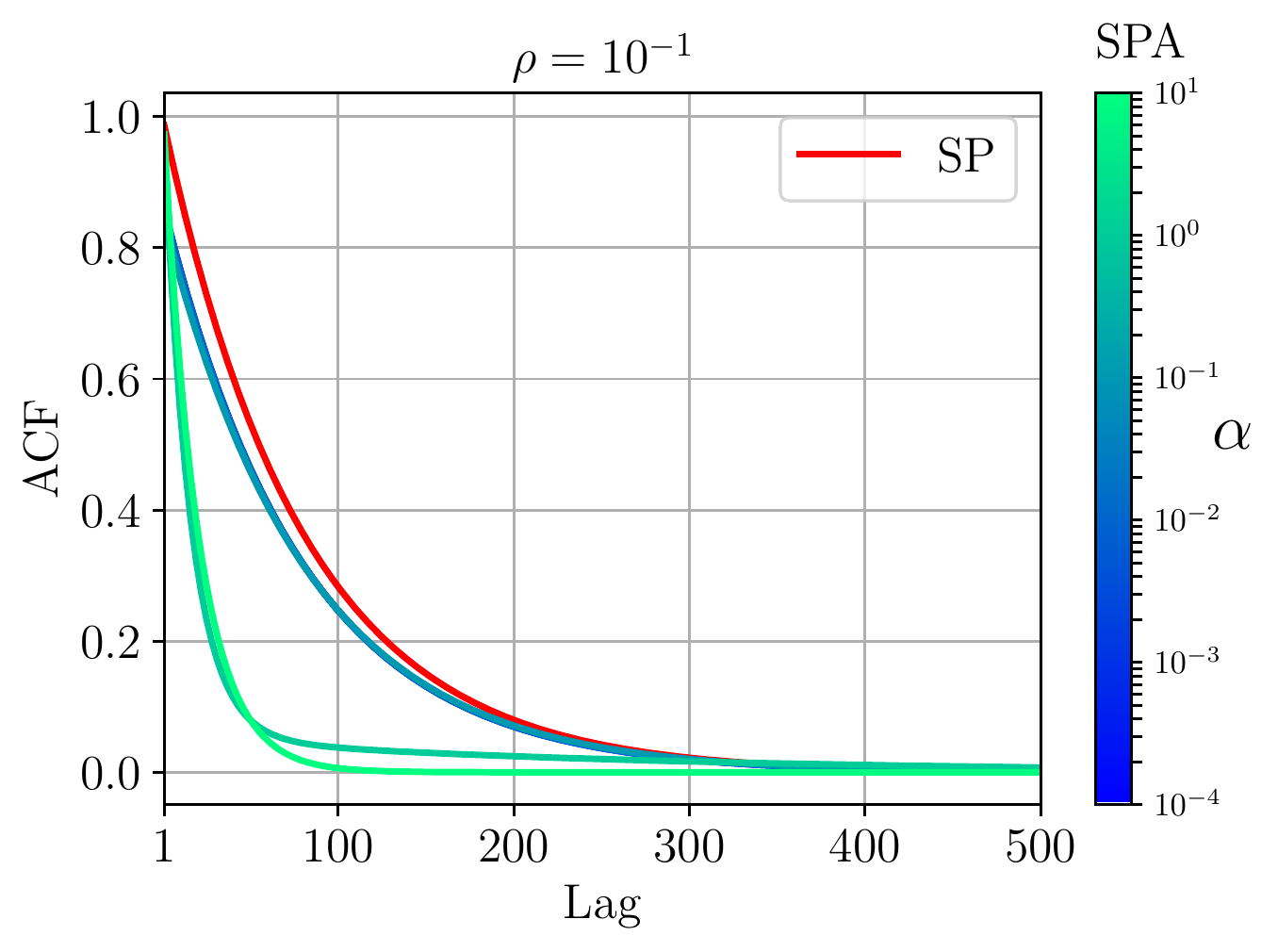}}}
  \mbox{{\includegraphics[scale=0.55]{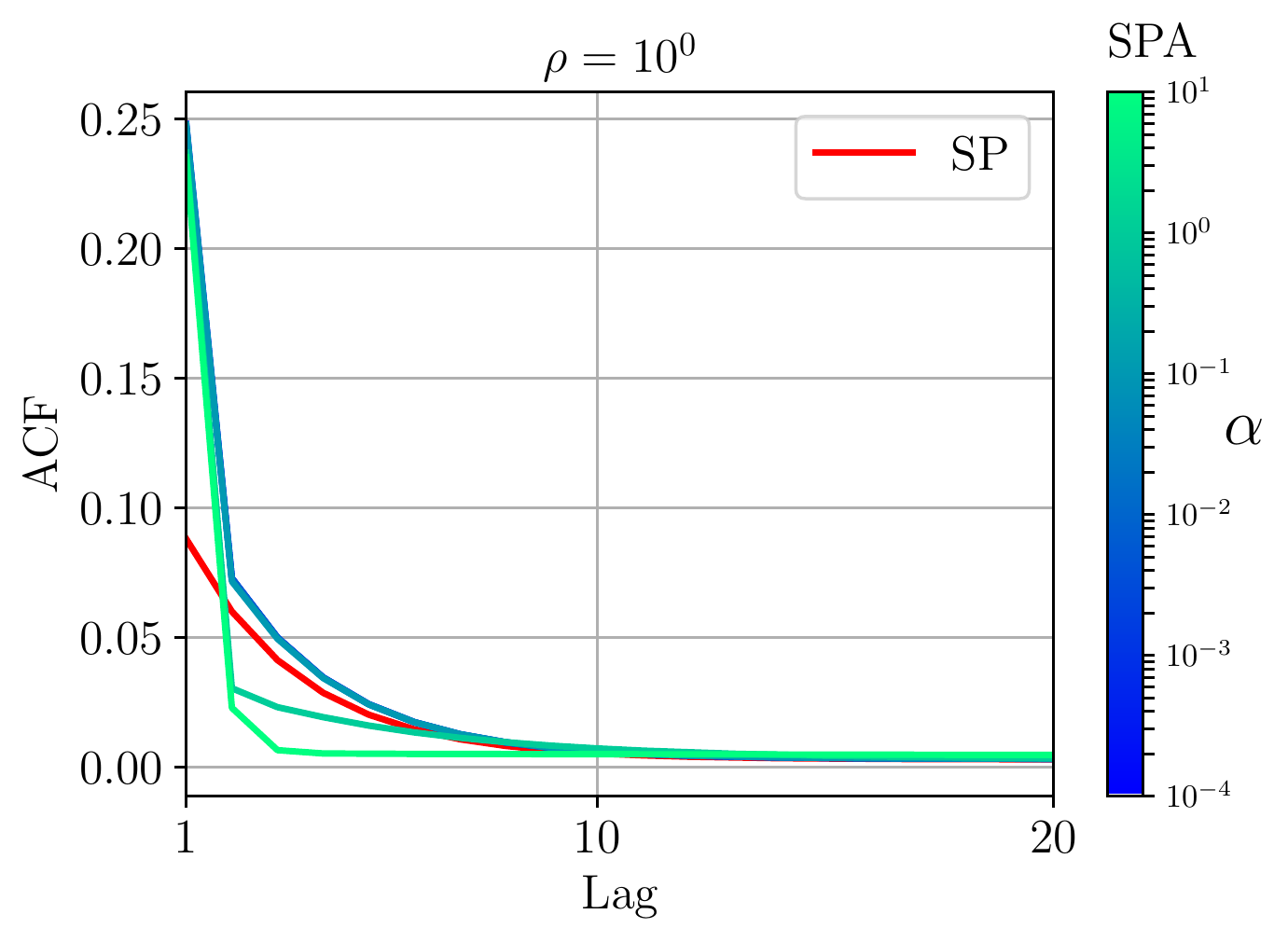}}}
  \caption{Image inpainting: effect of the parameter $\alpha$ (associated to the data augmentation step) for different values of the parameter $\rho$ on the autocorrelation functions of SPA (from \textcolor{guppiegreen}{guppie green} to \textcolor{blue}{blue}) and SP (\textcolor{red}{red}).
  The results were averaged over 10 independent runs.}
  \label{fig:ACF_alpha}
\end{figure}

Fig. \ref{fig:ACF_alpha} highlights the potential benefit of the data augmentation step described in \ref{subsec:data_augmentation}.
Thus, the autocorrelation functions associated to SP and SPA for different values of $\rho$ and $\alpha$ are depicted.
The latter were obtained by using $10^4$ samples and by considering the Markov chains from their first iteration (no burn-in period has been considered here).
The results are averaged over 10 independent runs.
The standard deviations being very small, they
are not depicted in Fig. \ref{fig:ACF_alpha}.
The effect of $\alpha$ for intermediate and large values of $\rho$ ($\rho \geq 1$ in this case) is not significant. However, as $\rho$ decreases, the impact of the data augmentation scheme governed by $\alpha$ on the autocorrelation function
increases significantly. This behavior is expected since this data augmentation was introduced to
bring an additional degree of freedom compared to the SP scheme when $\rho$ is small. Although the
limiting case $\rho \rightarrow 0$ is not considered in this experiment, it could be desired in some practical
scenarios. In such cases, considering the data augmentation step proposed in the manuscript can
bring a significant benefit concerning the exploration of the parameter space.

\subsubsection{Influence of $\rho$}

\begin{figure}
\centering
\includegraphics[width=3.5in]{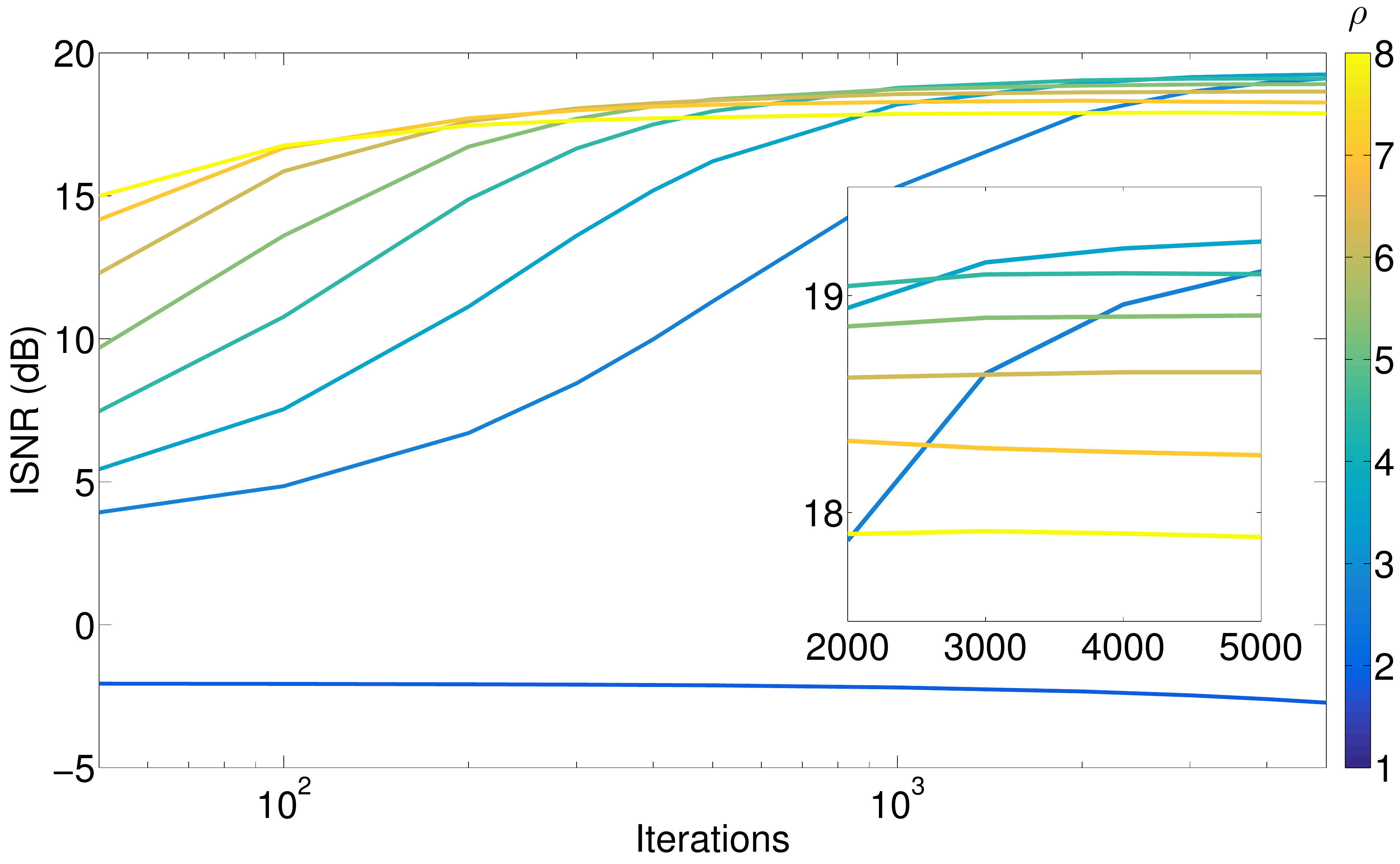}
\caption{Image inpainting: ISNR associated to SPA MMSE w.r.t. the number of iterations (in log-scale for the main figure and in normal-scale for the zoomed one) for different values of $\rho$.}
\label{fig:inpainting_ISNRvsETA}
\end{figure}

Fig. \ref{fig:inpainting_ISNRvsETA} shows the ISNR obtained with SPA on the Cameraman image w.r.t. the number of iterations and for different values of the parameter $\rho$ ranging from $\rho = 1$ (blue) to $\rho = 8$ (yellow).
High values of $\rho$ (yellow to green) rapidly lead to a stable but not optimal ISNR with low variance. Conversely, small values of $\rho$ (e.g. $\rho = 1$, dark blue) struggle to lead to an acceptable ISNR in a reasonable computational time.
On the other hand, intermediate values of $\rho$ (e.g. $\rho \in [2,4]$, blue to green) appear to be a trade-off between speed and precision of the estimation.
Thus, this range of values manages to lead, in a reasonable number of iterations, to an ISNR competing with the one obtained by SALSA (see Table \ref{table:inpainting_9_images}).

\subsubsection{Performance results}
Table \ref{table:inpainting_9_images} shows the average ISNR  obtained with the different algorithms for each image depicted in Fig. \ref{fig:9_images}.
P-MYULA applied to the original target distribution (\ref{eq:targetDistribution}) presents a lower ISNR on each image than the three other algorithms.
However, when P-MYULA is used within the SP or SPA frameworks, it manages to reach average performance similar to SALSA.
Note that the three MCMC approaches, contrary to the optimization algorithm SALSA, also carry credibility intervals for each pixel of the image to infer $\B{x}$, see Fig. \ref{fig:cameraman}(f).

\begin{table}
\renewcommand{\arraystretch}{1.3}
\caption{Image inpainting: average results over 25 different observation vectors in terms of ISNR for various algorithms and images. The ISNR associated to P-MYULA, SP and SPA was computed with the MMSE estimator.}
\label{table:inpainting_9_images}
\centering
\begin{tabular}{l c| c c c}
\thickhline
\bfseries & \bfseries SALSA & \bfseries P-MYULA& \bfseries SP& \bfseries SPA\\
\hline
Balloons & 26.18 & 23.00 & 26.19 & 26.18\\
Baboon & 14.37 & 13.35 & 14.60 & 14.59\\
Elaine & 23.61 & 21.21 & 23.86 & 23.84\\
Clock & 25.72 & 24.50 & 25.45 & 25.42\\
Donna & 24.71 & 21.69 & 23.87 & 23.82\\
House & 20.21 & 19.59 & 20.43 & 20.43\\
Peppers & 20.35 & 19.20 & 20.22 & 20.20\\
Cameraman & 19.48 & 18.76 & 19.34 & 19.34\\
Boat & 20.81 & 19.80 & 20.74 & 20.71\\
\thickhline
\end{tabular}
\end{table}

Table \ref{table:inpainting2} presents the numerical complexity resulting from one iteration along with the average number of iterations performed and the average computational time for each algorithm. The complexity of $O(N^2)$ refers to matrix-vector multiplication, that of $O(N)$ to the use of a proximal operator and $\mathcal{N}$ stands for the sampling from an univariate normal distribution. Note that the number of iterations and thereby the computational time of SALSA has been adapted to each observation to reach a target reconstruction error. This has not been the case for the MCMC algorithms where the total number of iterations has been fixed beforehand. Note that the cost of one MCMC iteration is roughly equivalent to the cost of one iteration in an ADMM framework. The difference in computational time is mainly related to the number of iterations performed by each algorithm. P-MYULA took on average roughly $3400$ longer time than SALSA.
Much more efficient, SP and SPA allowed to reduce the computing time w.r.t. P-MYULA by roughly $16$ by embedding P-MYULA and by simplifying its task.
This gain of computational time is mainly related to the Lipschitz constant of the gradient of the smooth potential used within P-MYULA.
Indeed, the convergence of P-MYULA, similarly to forward-backward splitting algorithms in optimization, is driven by the Lipschitz constant of the gradient of the smooth term in the potential $f + g$.
Namely, in this experiment, the Lipschitz constant $L_f$ of $\nabla f$ is given by $L_f = \sigma^{-2}\lambda_{\mathrm{max}}(\B{H}^T\B{H})$, where $\lambda_{\mathrm{max}}(\B{H}^T\B{H})$ is the largest eigenvalue of $\B{H}^T\B{H}$.
This constant is highly dependent on the problem, more precisely on the forward operator $\B{H}$ and cannot be tuned.
On the contrary, if the proposed variable splitting approach is used, P-MYULA is now embedded in the Gibbs sampling scheme and is used to sample from \eqref{eq:inpainting_condDistrib_z}.
In \eqref{eq:inpainting_condDistrib_z}, the relevant Lipschitz constant is $L_f' = \rho^{-2}$: this constant now can be chosen carefully to improve the mixing and accelerate the convergence of P-MYULA within SPA, see Fig. \eqref{fig:inpainting_ISNRvsETA}.

\begin{table}
\renewcommand{\arraystretch}{1.3}
\caption{Image inpainting: computational complexity related to one iteration, average number of iterations performed and average computational time for each algorithm.}
\label{table:inpainting2}
\centering
\begin{tabular}{l l l l}
\thickhline
\bfseries & \bfseries computational complexity & \bfseries $\#$ iterations & \bfseries time (s) \\
\hline
SALSA & $O(N^2) + O(N)$ & 43 & 1\\ \hline
P-MYULA & $O(N^2) + O(N) + N \mathcal{N}$ & $10^5$ & 3408\\
SP & $O(N^2) + O(N) + 3N \mathcal{N}$ & $5 \times 10^3$ & 207\\
SPA & $O(N^2) + O(N) + 4N \mathcal{N}$& $5 \times 10^3$ & 215\\
\thickhline
\end{tabular}
\end{table}

Fig. \ref{fig:cameraman} shows the results obtained by SPA on the Cameraman image. Those obtained by SP were similar and are omitted here for brevity. The MMSE estimators of $\B{x}$ and $\B{z}$ are very close, ensuring that the proposed variable splitting method behaves successfully.
The variable splitting residuals contained in $\B{u}$ appear to be close to $0$ for most pixels but present a certain structure.
Thus, positive and negative residuals seem to share a complementary structure near the boundaries of objects in the image. This particular structure of the residuals is confirmed by the analysis of the credibility intervals: there is more uncertainty (of about $80$ grey-levels) on the object contours of the image. The same conclusion was drawn in \cite{Pereyra2016B} when P-MALA was applied to an image deblurring problem with total variation.

\begin{figure}[!t]
\centering
\includegraphics[width=3.5in]{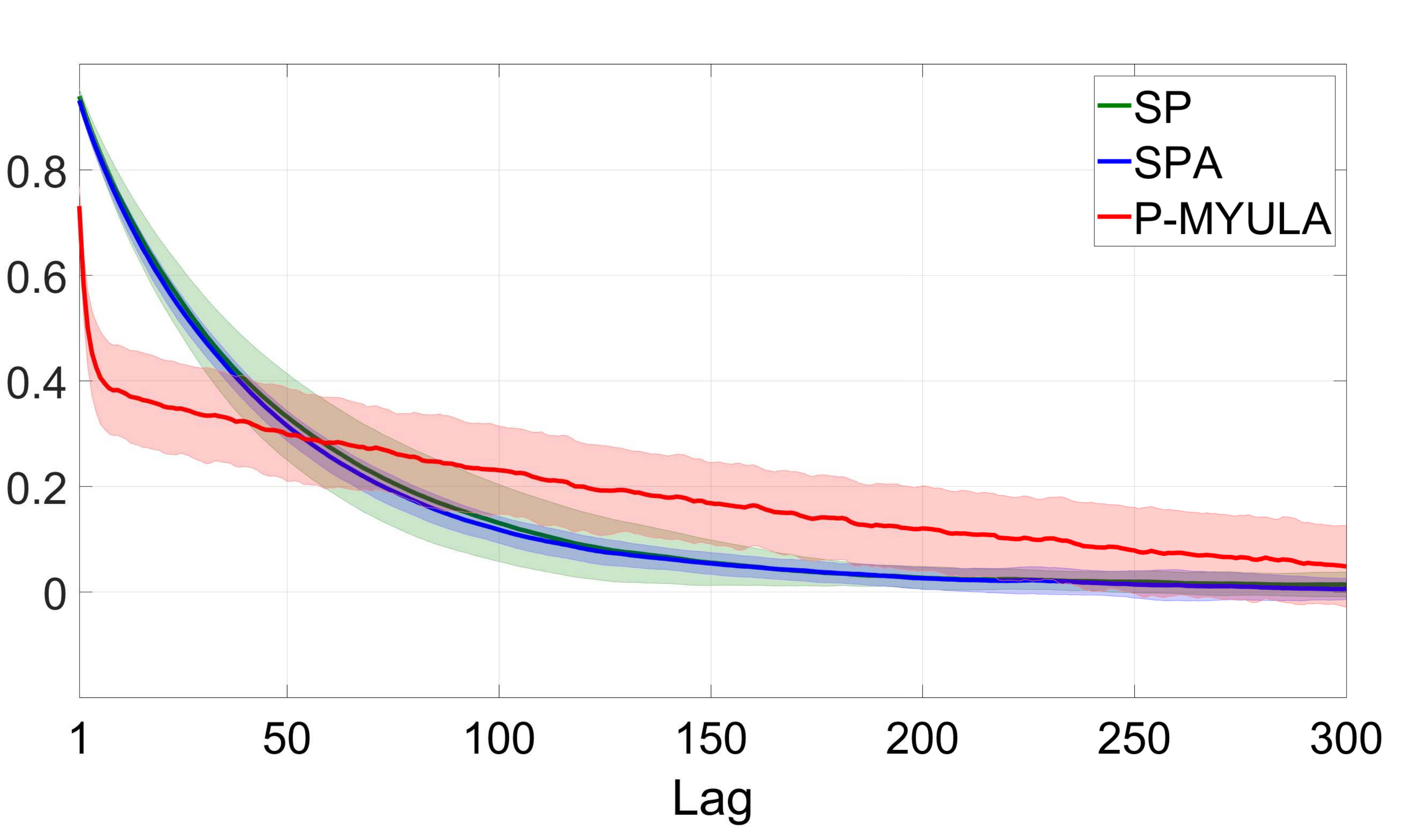}
\caption{Image inpainting: average chain autocorrelation functions of SP (green), SPA (blue) and P-MYULA (red).
Shaded areas represent the intervals corresponding to the standard deviation computed over 25 trials.}
\label{fig:inpainting_ACF}
\end{figure}

Fig. \ref{fig:inpainting_ACF} compares the average autocorrelation functions (using $-\log p(\B{x}|\B{y})$ as a scalar summary and obtained after the burn-in period) of SP, SPA and P-MYULA on the Cameraman image.
The shaded regions depicted in Fig. \ref{fig:inpainting_ACF} represent the standard deviation ranges associated to each MCMC algorithm.
SP and SPA present better mixing properties than P-MYULA,
showing that the proposed approaches successfully and more efficiently explore their respective parameter space.
Additionally, although the average autocorrelation functions of SP and SPA are similar, the data augmentation scheme within SPA led to a Markov chain with more stable mixing properties over different observations (see the green and blue shaded areas).
Note that the potential benefit of the data augmentation step detailed in Section \ref{subsec:data_augmentation} increases when $\rho$ decreases.

\subsubsection{Discussion}
The expectations from MCMC algorithms like SP, SPA and P-MYULA are threefold.
Firstly, to infer the hidden image $\B{x}$, the MCMC methods are expected to efficiently explore the parameter space, in particular nearby the high potential regions. Secondly, the computational cost of these algorithms should remain reasonable compared to SALSA.
Finally, they have to produce Markov chains with good mixing properties in order to explore the entire probability distribution and thus provide accurate credibility intervals.

Based on the previous results, SP and SPA appear as a very good trade-off between these three expectations: mixing properties, efficient exploration and reasonable computational cost.
The latter expectation is particulary satisfied. Yet, even though the computing times associated to the proposed approaches are reasonable, they are roughly $200$ times higher than SALSA for a problem in high dimension ($N = 65536$). This overhead cost results from the exploration of the parameter space: this is the price to pay to derive confidence intervals on the inferred parameter, and it seems difficult to get cheaper methods. 

\section{Conclusion}
\label{sec:Conclusion}
This paper introduced a new general Bayesian framework which aims at solving large-scale inference problems.
To derive the proposed methodology, two new optimization-driven hierarchical Bayesian models and their associated MCMC algorithms, inspired from variable splitting and data augmentation, were introduced.
Similarly to the ADMM in an optimization context, the proposed approach could be summarized as a ``divide and conquer" method.
Thus, the derived algorithms lead to simpler sampling steps so that efficient state-of-the-art MCMC algorithms can be embedded for each sampling task.
Note that the proposed approach can also be used to distribute MCMC methods on multiples machines as detailed in \cite{Rendell2018}.

The versatility and efficiency of the proposed algorithms have been assessed on two often-studied problems and compared to recent state-of-the-art optimization and sampling approaches.
Based on these results, SP and SPA appear to be more efficient while sharing a large scope of applications.
Additionally, their reasonably low computational cost compared to optimization algorithms helps to reduce the gap between optimization and simulation-based approaches while providing precious credibility intervals.

Future works will focus on other forms for the functions $f$, $g$, $\phi_1$ and $\phi_2$ to illustrate the broad scope of applications of the proposed approach.
In particular, it will include inference problems involving non-convex target distributions.
Finally, this paper presented SP and SPA as efficient algorithms designed to solve an inference problem.
They could also be used to approximate complex target distributions.
In this approximation context, future works will include a theoretical analysis of the proposed approach. 

\appendices

\section{Proof of Theorem \ref{theorem:1}}
\label{appendice:proof_theorem1}
\begin{IEEEproof}
  The usual target distribution (\ref{eq:targetDistribution}) has the form
\begin{align}
    \pi(\B{x}) = \dfrac{\exp\br{-f(\B{x})-g(\B{x})}}{\int_{\mathbb{R}^N} \exp\br{-f(\B{x})-g(\B{x})}\mathrm{d}\B{x}}, \label{posterior2}
\end{align}
and has been assumed to define a proper probability distribution.
By denoting
\begin{align}
  p_{\phi_1}(\B{x},\B{z};\rho^2) \triangleq \dfrac{\exp\br{- \phi_1(\B{x},\B{z};\rho^2)}}{\int_{\mathbb{R}^N}\exp\br{- \phi_1(\B{x},\B{z};\rho^2)}\mathrm{d}\B{z}},
\end{align}
the split-distribution \eqref{eq:jointDistribution} writes
\begin{align}
    \pi_{\rho}(\B{x},\B{z}) = \dfrac{\exp\br{-f(\B{x})-g(\B{z})}p_{\phi_1}(\B{x},\B{z};\rho^2)}{\int_{\mathbb{R}^N}\int_{\mathbb{R}^N} \exp\br{-f(\B{x})-g(\B{z})}p_{\phi_1}(\B{x},\B{z};\rho^2)\mathrm{d}\B{z}\mathrm{d}\B{x}}. \label{posteriorSplit2}
\end{align}
Let define
\begin{align}
    p_{\rho}(\B{x}) = \int_{\mathbb{R}^N}\pi_{\rho}(\B{x},\B{z})\mathrm{d}\B{z}. \label{marginalPosteriorSplit2}
\end{align}
Under the two distributions (\ref{posterior2}) and (\ref{marginalPosteriorSplit2}), we are interested in showing that
\begin{align}
    \nr{\pi - p_{\rho}}_{\mathrm{TV}} &= \int_{\mathbb{R}^N}\abs{\pi(\B{x}) - p_{\rho}(\B{x})}\mathrm{d}\B{x}
\end{align}
tends towards zero when $\rho^2 \rightarrow 0$.

Assumption \ref{assumption:1} implies that
\begin{align}
    &\lim_{\rho^2\to0} \exp\br{-f(\B{x})-g(\B{z})}p_{\phi_1}(\B{x},\B{z};\rho^2) \nonumber \\
    &= \exp\br{-f(\B{x})-g(\B{z})}\delta_{\B{x}}(\B{z}).\label{convNumerator}
\end{align}

Since $\forall \rho>0$, $\exp\br{-f(\B{x})-g(\B{z})}p_{\phi_1}(\B{x},\B{z};\rho^2)$ has been supposed to be integrable, see Section \ref{subsec:variable_splitting}, it follows from the dominated convergence theorem that
\begin{align}
    \lim_{\rho^2\to0}&\int_{\mathbb{R}^N}\int_{\mathbb{R}^N} \exp\br{-f(\B{x})-g(\B{z})}p_{\phi_1}(\B{x},\B{z};\rho^2)\mathrm{d}\B{z}\mathrm{d}\B{x} \\
    &=\int_{\mathbb{R}^N}\int_{\mathbb{R}^N}\exp\br{-f(\B{x})-g(\B{z})}\delta_{\B{x}}(\B{z})\mathrm{d}\B{z}\mathrm{d}\B{x} \label{convDenominator2}\\
    &=\int_{\mathbb{R}^N}\exp\br{-f(\B{x})-g(\B{x})}\mathrm{d}\B{x}. \label{convDenominator}
\end{align}

Combining (\ref{convNumerator}) and (\ref{convDenominator}), it follows
\begin{align}
    \lim_{\rho^2\to0} \pi_{\rho}(\B{x},\B{z}) = \dfrac{\exp\br{-f(\B{x})-g(\B{z})}\delta_{\B{x}}(\B{z})}{\int_{\mathbb{R}^N}\exp\br{-f(\B{x})-g(\B{x})}\mathrm{d}\B{x}}.\label{conv_p_rho}
\end{align}
Using one more time the dominated convergence theorem, as in \eqref{convDenominator2} and \eqref{conv_p_rho} leads for all $\B{x} \in \mathbb{R}^N$ to
\begin{align}
    \lim_{\rho^2\to0} p_{\rho}(\B{x}) = \dfrac{\exp\br{-f(\B{x})-g(\B{x})}}{\int_{\mathbb{R}^N}\exp\br{-f(\B{x})-g(\B{x})}\mathrm{d}\B{x}} = \pi(\B{x}).\label{conv_pi_rho}
\end{align}

Finally, Scheff\'e's lemma \cite{Scheffe1947} ensures the convergence of $p_{\rho}$ towards $\pi$ in total variation, that is
\begin{align}
    \lim_{\rho^2\to0}\nr{\pi - p_{\rho}}_{\mathrm{TV}} = \lim_{\rho^2\to0} \int_{\mathbb{R}^N}\abs{\pi(\B{x}) - p_{\rho}(\B{x})}\mathrm{d}\B{x} = 0.
\end{align}
\end{IEEEproof}

\section{Case of multiple functions $h_i$}
\label{appendice:generalized_SP_SPA}
Assume that the problem considered involves the introduction of $N_h$ functions $h_i$ along with $N_h$ observation operators $\B{K}_i \in \mathbb{R}^{k_i \times N}$, $i \in \{1,\dots,N_h\}$.
Thereby, the usual target distribution takes the form
\begin{align}
  \pi(\B{x}) \propto \exp\br{-\sum_{i=1}^{N_h}h_i(\B{K}_i\B{x})}.
\end{align}
\begin{remark}
  In the case where $N_h = 2$ and $\B{K}_1 = \B{K}_2 = \B{I}_N$, the usual target distribution defined in (\ref{eq:targetDistribution}) is retrieved.
\end{remark}
\subsection{Derivation of SP}
In order to simplify the sampling procedure, let introduce $N_h$ splitting variables denoted $\B{z}_1,\B{z}_2,\ldots,\B{z}_{N_h} \in \mathbb{R}^{k_i}$, a positive parameter $\rho$ and $N_h$ divergences $\phi_i$ defined on $\mathbb{R}^{k_i} \times \mathbb{R}^{k_i}$ such that the underlying joint probability distribution has the form
\begin{align}
  p(\B{x},\B{z}_1,\B{z}_2,\ldots,\B{z}_{N_h};\rho^2) &\propto \exp\br{-\sum_{i=1}^{N_h}h_{i}(\B{z}_i) \right. \nonumber \\
  &\left. + \phi_i\pr{\B{K}_i\B{x},\B{z}_i;\rho^2}}.
\end{align}
Thereby, the generalized SP implies the sampling from the conditional distributions
\begin{align}
  &p(\B{x}|\B{z}_{i,i \in \{1,\ldots,N_h\}};\rho^2) \propto \exp\br{-\sum_{i=1}^{N_h}\phi_i\pr{\B{K}_i\B{x},\B{z}_i;\rho^2}}, \\
  &p(\B{z}_i|\B{x};\rho^2) \propto \exp\br{-h_i(\B{z}_i) -\phi_i\pr{\B{K}_i\B{x},\B{z}_i;\rho^2}},
\end{align}
for all $i \in \{1,\ldots,N_h\}$.

\subsection{Derivation of SPA}
In the same manner, let introduce $N_h$ splitting and auxiliary variables denoted $\B{z}_1,\B{z}_2,\ldots,\B{z}_{N_h} \in \mathbb{R}^{k_i}$ and $\B{u}_1,\B{u}_2,\ldots,\B{u}_{N_h} \in \mathbb{R}^{k_i}$, respectively.
Additionally, let introduce positive parameters $\rho$ and $\alpha$, $N_h$ divergences $\phi_i$ defined on $\mathbb{R}^{k_i} \times \mathbb{R}^{k_i}$ and $N_h$ functions $\psi_i$ defined on $\mathbb{R}^{k_i}$ such that the underlying joint probability distribution has the form
\begin{align}
  &p(\B{x},\B{z}_{i,i \in \{1,\ldots,N_h\}},\B{u}_{i,i \in \{1,\ldots,N_h\}};\rho^2,\alpha^2) \propto \nonumber \\
  &\exp\br{-\sum_{i=1}^{N_h}h_{i}(\B{z}_i) + \phi_i\pr{\B{K}_i\B{x},\B{z}_i-\B{u}_i;\rho^2} + \psi_i(\B{u}_i;\alpha^2)}.
\end{align}
The generalized SPA implies the sampling from the conditional distributions
\begin{align}
  &p(\B{x}|\B{z}_i,\B{u}_i;\rho^2) \propto \exp\br{- \sum_{i=1}^{N_h}\phi_i\pr{\B{K}_i\B{x},\B{z}_i - \B{u}_i;\rho^2}}, \\
  &p(\B{z}_i|\B{x},\B{u}_i;\rho^2) \propto \exp\br{-h_i(\B{z}_i) - \phi_i\pr{\B{K}_i\B{x},\B{z}_i-\B{u}_i;\rho^2}},
\end{align}
for all $i \in \{1,\ldots,N_h\}$, and
\begin{align}
  &p(\B{u}_i|\B{x},\B{z}_i;\rho^2,\alpha^2) \propto \exp\br{-\psi_i(\B{u}_i;\alpha^2) \right. \\ \nonumber
  & \hspace{2.5cm} \left. - \phi_i\pr{\B{K}_i\B{x},\B{z}_i-\B{u}_i;\rho^2}},
\end{align}
for all $i \in \{1,\ldots,N_h\}$.

\section{Efficient Gaussian sampling in high dimension}
\label{appendice:efficientGaussianSampling}

In this Appendix, notations are those of Section \ref{subsec:BayesProblems_Gaussian}.
Suppose that one wants to sample efficiently from the high-dimensional Gaussian conditional distributions
  \begin{align}
    &p(\B{z}|\B{x},\B{u}) = \mathcal{N}\pr{\B{m_z},\B{G_z}^{-1}} \label{eq:gaussian_condDistrib_z} \\
    &p(\B{x}|\B{z},\B{u}) = \mathcal{N}\pr{\B{m_x},\B{G_x}^{-1}} \label{eq:gaussian_condDistrib_x}
  \end{align}
where, in particular,
\begin{numcases}{}
\B{G_z} = \gamma\B{L}^T\B{L} + \dfrac{1}{\rho^2}\B{I}_N. \label{eq:gaussianPosterior_cov_z} \\
\B{G_x} = \B{H}^T\boldsymbol{\Omega}\B{H} + \dfrac{1}{\rho^2}\B{I}_N \label{eq:gaussianPosterior_cov_x}
\end{numcases}

\subsection{Efficient sampling from (\ref{eq:gaussian_condDistrib_z})}
\label{appendice:efficientGaussianSampling_z}
The matrix $\B{L}$ was assumed to be a circulant matrix.
Thereby, the latter can be diagonalized in the Fourier domain such that
\begin{align}
  \B{L} = \B{F}^H\B{\Lambda_L}\B{F}, \label{eq:Fourierdecomp_L}
\end{align}
where $\B{F}$ and $\B{F}^H$ are unitary matrices ($\B{F}^H\B{F} = \B{F}\B{F}^H = \B{I}_N$) associated with the Fourier and inverse Fourier transforms.
$\B{\Lambda_L}$ is the diagonal counterpart of $\B{L}$ in the Fourier domain.
Using (\ref{eq:Fourierdecomp_L}), the precision matrix defined in (\ref{eq:gaussianPosterior_cov_z}) has the form
\begin{align}
  \B{G_z} &= \gamma\B{F}^H\B{\Lambda_L}^H\B{F}\B{F}^H\B{\Lambda_L}\B{F} + \dfrac{1}{\rho^2}\B{I}_N \nonumber \\
  &= \gamma\B{F}^H\B{\Lambda_L}^H\B{\Lambda_L}\B{F} + \dfrac{1}{\rho^2}\B{I}_N
\end{align}
Then, the counterpart of $\B{G_z}$ in the Fourier domain is diagonal and has the form
\begin{align}
  \B{\Lambda_{G_z}} &= \gamma\B{\Lambda_L}^H\B{\Lambda_L} + \dfrac{1}{\rho^2}\B{I}_N. \label{eq:Fourierdecomp_Gz}
\end{align}
Using (\ref{eq:Fourierdecomp_Gz}), one can efficiently sample from (\ref{eq:gaussian_condDistrib_z}) by drawing $N$ independent Gaussian samples in the Fourier domain.

\subsection{Efficient sampling from (\ref{eq:gaussian_condDistrib_x})}
\label{appendice:efficientGaussianSampling_x}
Unfortunately, although the matrix $\B{H}$ was assumed circulant, the first term in (\ref{eq:gaussianPosterior_cov_x}) cannot be diagonalized in the Fourier domain.
To cope with this problem, the auxiliary method of \cite{Marnissi2018} is used.
An additional variable $\B{v} \in \mathbb{R}^N$ is introduced such that the conditional distributions of $\B{x}$ and $\B{v}$ are
\begin{align}
    &p(\B{x}|\B{z},\B{u},\B{v}) = \mathcal{N}\pr{\B{\tilde{m}_x},\B{\tilde{G}_x}^{-1}} \label{eq:Fourierdecomp_x}\\
    &p(\B{v}|\B{x}) = \mathcal{N}\pr{\B{m_v},\B{G_v}^{-1}} \label{eq:Fourierdecomp_v}
  \end{align}
where, in particular,
\begin{numcases}{}
\B{\tilde{G}_x} = \dfrac{1}{\mu_1}\B{H}^T\B{H} + \dfrac{1}{\rho^2}\B{I}_N \label{eq:gaussianPosterior_cov_xx}\\
\B{G_v}^{-1} = \dfrac{1}{\mu_1}\B{I}_N - \B{\Omega}. \label{eq:gaussianPosterior_cov_v}
\end{numcases}
\begin{remark}
  The positive parameter $\mu_1$ is such that $\mu_1\nr{\B{\Omega}}_S < 1$ ($\nr{.}_S$ stands for the spectral norm of a matrix) ensuring that (\ref{eq:gaussianPosterior_cov_v}) is positive definite.
\end{remark}

As in Appendix \ref{appendice:efficientGaussianSampling_x}, the matrix $\B{H}$ (assumed circulant) can be diagonalized in the Fourier domain.
Under these two conditional distributions, $\B{x}$ can be efficiently drawn in the Fourier domain and $\B{v}$ can be efficiently sampled in $\mathbb{R}^N$ as $\B{\Omega}$ was assumed diagonal. 

\ifCLASSOPTIONcaptionsoff
  \newpage
\fi

\bibliographystyle{IEEEtran}
\bibliography{IEEEabrv,bibliocleanIEEE_nd}

\end{document}